\documentclass[11pt,a4]{article}

\usepackage{amsmath,amssymb}
\usepackage{graphicx}
\usepackage{indentfirst}
\usepackage{csquotes}
\usepackage{algorithm}
\usepackage{algorithmicx}
\usepackage{algpseudocode}
\usepackage{booktabs}
\usepackage{multirow}
\usepackage{bm}
\usepackage{subcaption}
\usepackage{tikz}
\usepackage{pgfplots}
\pgfplotsset{compat=1.18}
\usepackage{textcomp}
\usepackage{comment}
\usepackage{rotating}
\usepackage{footmisc}
\usepackage{caption}

\usepackage{xcolor}       
\usepackage{colortbl}     
\definecolor{myorange}{rgb}{1.0, 0.5, 0.0} 
\usepackage{hyperref}
\hypersetup{
    colorlinks=true,
    linkcolor=blue,
    filecolor=blue,
    urlcolor=blue,
    citecolor=blue
}

\newtheorem{remark}{Remark}

\newcommand{\bfx}{{\mathbf{x}}}
\newcommand{\bfy}{{\mathbf{y}}}
\newcommand{\bfY}{{\mathbf{Y}}}
\newcommand{\bfW}{{\mathbf{W}}}
\newcommand{\bfZ}{{\mathbf{Z}}}
\newcommand{\bfU}{{\mathbf{U}}}

\newcommand{\bfA}{{\mathbf{A}}}
\newcommand{\bfI}{{\mathbf{I}}}
\newcommand{\bfV}{{\mathbf{V}}}
\newcommand{\bmV}{{\bm{V}}}

\newcommand{\bfw}{{\mathbf{w}}}
\newcommand{\bfz}{{\mathbf{z}}}
\newcommand{\bmSigma}{{\bm{\Sigma}}}
\newcommand{\bmLambda}{{\bm{\Lambda}}}
\newcommand{\bfb}{{\mathbf{b}}}

\newcommand{\bfT}{{\mathbf{T}}}

\begin{document}

\title{Extracting transient Koopman modes from short-term weather simulations with sparsity-promoting dynamic mode decomposition}  
\date{\today}

\author{
  Zhicheng Zhang\thanks{Department of Electrical, Electronic and Digital Science and Engineering, Kyoto University, Katsura, Kyoto 615-8510, Japan.
  (e-mail: \text{zhang.zhicheng.2c@kyoto-u.ac.jp}; 
  \text{susuki.yoshihiko.5c@kyoto-u.ac.jp})}
  \and Yoshihiko Susuki\footnotemark[1] 
  \and \quad Atsushi Okazaki\thanks{Institute of Advanced Academic Research, Center for Environmental Remote Sensing, Chiba University, 1-33 Yayoi-cho, Inage, Chiba 263-0022, Japan.
  (e-mail: \text{atsushi.okazaki@chiba-u.jp})}
}


\maketitle

\begin{abstract} 
Convective features, represented here as warm bubble-like patterns, reveal essential high-level information about how short-term weather dynamics evolve within a high-dimensional state space. In this paper, we introduce a data-driven framework that uncovers transient dynamics captured by Koopman modes responsible for these structures and traces their emergence, growth, and decay. 
Our approach applies the sparsity-promoting dynamic mode decomposition to weather simulations, yielding a few number of selected modes whose sparse amplitudes highlight dominant transient structures. By tuning the sparsity weight, we balance reconstruction accuracy and model complexity. We illustrate the methodology on weather simulations, using the magnitude of velocity and vorticity fields as distinct observable datasets. 
The resulting sparse dominant Koopman modes capture the transient evolution of bubble-like pattern and can reduce the dimensionality of weather system model, offering an efficient surrogate for diagnostic and forecasting tasks.
\end{abstract}


\maketitle 
\section{Introduction}
Modeling weather or climate systems poses a rigorous challenge due to the inherent complexity and variability of real-world meteorological phenomena \cite{hartmann2015global}. 
In contrast to simple mathematical models, such as the classic Lorenz system or the Navier-Stokes equations, which are commonly formulated in geophysical fluid dynamics and driven by ordinary or partial differential equations, real-world weather dynamics involve high-dimensional (or even infinite-dimensional) nonlinear systems \cite{pedlosky2013geophysical}. 
These complexities go far beyond the scope of traditional chaotic or turbulent behaviors, requiring more practical, analytical, and computational approaches. 
For example, a success story in meteorology is weather prediction, which has been greatly enhanced through the integration of better theoretical models, improved computational capabilities, and advanced observational data-enabled systems. 
These advancements allow for short-term weather forecasting, as accurate as possible, and data assimilation into weather or climate modeling \cite{kotsuki2020data}.

Traditional climate or weather analysis is based on the empirical orthogonal function (EOF) method \cite{lorenz1956empirical}, which is similar to principal component analysis and effectively identifies coherent structures by decomposing spatiotemporal data into orthogonal spatial patterns and their corresponding temporal coefficients of climate variability \cite{hannachi2007empirical}.
Extracting leading modes enables dimension reduction in large-scale geospatial climate and weather data, with applications such as investigating sea-level pressure anomalies, sea surface temperature (SST), and sea ice variability \cite{schmidt2019spectral}. 
To enhance both accuracy and scalability, frequency-domain extensions of the EOF method, such as singular spectrum analysis and nonlinear Laplacian spectral analysis, have been incorporated into climatological studies \cite{giannakis2012nonlinear}.

Despite the existing fruitful results in earth system models for climate enumerations, these methods often assume that measured meteorological time-series data are sufficiently rich (or data informativity), providing a \emph{long-term} snapshots that primarily capture the \emph{steady-state} dynamics of climate or weather systems. 
On the one hand, in practice, data-driven analysis of weather systems is often constrained by the availability of finite-length time series, corresponding to short-term snapshots. From an observational perspective, dynamical features that dominate only over a finite observation window are naturally interpreted as \emph{transient}.
On the other hand, from a dynamical systems perspective, many geophysical and engineering systems intrinsically exhibit transient behavior, such as transient stability in power systems \cite{chiang2011direct}, non-normality in hydrodynamics \cite{trefethen1993hydrodynamic,holmes1996turbulence}, and perturbation growth in geophysical fluid dynamics \cite{pedlosky2013geophysical}.
In this work, we focus on the former, finite-length data-driven perspective on transient weather dynamics, in which transient behaviors are defined as short-term, time-localized features that dominate the observed evolution over finite time horizons, with their apparent importance governed by the available observation window and measurement process.

The price of some strategies, such as wavelet analysis and empirical orthogonal functions (EOFs), lack a clear connection to the underlying dynamics of the field on which they are computed. In particular, they are not directly related to linear instability modes \cite{dommenget2002cautionary} and are not suitable for capturing the system’s evolution or for providing a meaningful simplification of its dynamics, as also discussed in \cite{navarra2024variability}.
Recent trends in statistical methods, such as linear inverse models (LIM) \cite{kwasniok2022linear} and non-parametric model-analog techniques (MAT), have been employed to explore climate changes and improve forecasting for phenomena such as the El Ni\~{n}o–Southern Oscillation (ENSO) \cite{wang2024role}.  However, these methods face limitations: LIM is restricted to linear dynamics and cannot capture nonlinear interactions, even though it is useful for medium-term forecasts and understanding low-frequency variability; MAT, in general, requires rich historical data and struggles to generalize to high-dimensional systems.

{Koopman operator} offers a novel operator-theoretic perspective for analyzing nonlinear dynamical systems with linear operator paradigm~\textcolor{blue}{\cite{mezic2005spectral,mauroy2020koopman}}. By lifting dynamics from a finite-dimensional state space to an infinite-dimensional space of observables, the Koopman framework yields an exact, global, linear representation of nonlinear evolution \cite{mezic2005spectral,rowley2009spectral}. Its spectral analysis exposes dominant spatial and temporal structures, achieving model reduction. This is the essence of \emph{Koopman mode decomposition} (KMD), in which nonlinear or multivariate time series can be decomposed into an infinite sum of oscillatory modes with single frequencies, called the \emph{Koopman modes}; see, e.g., \cite{rowley2009spectral,schmid2010dynamic,mezic2013analysis}.  
The oscillatory features are characterized by the eigenvalues of the Koopman operator, called the \emph{Koopman eigenvalues}. 
Prior to the advent of data-assistant approaches, Galerkin methods and proper orthogonal decomposition (POD) \cite{holmes1996turbulence} were the primary computational techniques for modal analysis and mode decomposition of nonlinear flows, which also enables reduced-order models via the projection operators.
Over the past decade, \emph{dynamic mode decomposition} (DMD), the-state-of-the-art modern data-driven algorithm, has demonstrated remarkable performance in approximating the KMD directly from practical or experimental data; numerically, DMD performs an Arnoldi-type procedure on the snapshot matrix \cite{rowley2009spectral,schmid2010dynamic}. 
Moreover, rigorous theoretical connections and guarantees between DMD and KMD have been well established, see, e.g., in~\cite{tu2014dynamic}.
Consequently, numerous variants of DMD (see, e.g., \cite{jovanovic2014sparsity,williams2015data,kutz2016dynamic}) and the framework of applied Koopmanism \cite{budivsic2012applied} for data-driven dynamics modelings have been successfully deployed in a wide range of disciplines. For example, these include fluid dynamics \cite{schmid2010dynamic,mezic2013analysis}, power and energy systems \cite{susuki2013nonlinear}, 
as well as recent advances in machine learning and deep learning \cite{takeishi2017learning,lusch2018deep,yeung2019learning}.
For comprehensive coverage of DMD and KMD methodologies, see textbooks \cite{kutz2016dynamic,brunton2022data} and \cite{mauroy2020koopman}.
In Section~\ref{sec:literature-rew}, we provide a focused literature review of Koopman operator applications in weather and climate systems.
%

\subsection{Contents and Challenges: Short-Term Transient Weather Dynamics}
In this paper, we will focus on a critical question: \emph{How can observed short-term snapshots be used to construct a low-dimensional representation of inherently transient weather dynamics}? 
Namely, how can we extract only a small number of modes to represent transient weather dynamics from finite-length data?
This challenge becomes particularly striking when only a finite-length data at hand, as it is driven by \emph{transient dynamics} before the weather system either stabilizes or evolves in a non-stationary manner. 
This distinction clarifies the fundamental difference between climate and weather systems. Prior work has primarily analyzed long-term data collections (e.g., SST time series) associated with \emph{steady-state} climate dynamics, rather than \emph{short-term, finite-length} datasets related to transient weather dynamics (see Section~\ref{sec:literature-rew}). By leveraging Koopman theory and dynamic mode decomposition (DMD), these studies identify the leading oscillatory modes of climate variability (e.g.,~\cite{navarra2024variability,navarra2021estimation,womack2026theoretical,zhang2025sparsity}). This, in turn, motivates us to adopt KMD and DMD within a data-driven framework to investigate transient weather dynamics from short-term data, with a focus on extracting the associated dominant modes.
In this paper, we use weather simulations conducted with the Scalable Computing for Advanced Library Environment (SCALE)~\cite{nishizawa2020scale} to collect weather data, as shown in Fig.~\ref {fig:SCALE}. The simulations focus on warm bubble-like experiments under various atmospheric factors, including ``velocity" and ``vorticity" magnitudes, ``temperature," ``humidity ratio," and ``pressure" fields from SCALE data, which are referred to as \emph{observables}. Notably, this experiment is based on real-world convective scenarios for precipitation evaluation \cite{ohtsuka2024convex} and lasts only 1-2 hours, representing a short-time weather evolution governed by transient dynamics in a finite-length dataset.
From the physical implications, such warm bubble-like patterns are often analyzed within convection-permitting weather models \cite{honda2025exploring}, where rising warm air from cumulonimbus clouds interacts with descending cumulus clouds. 
These structures play a pivotal role in weather and atmospheric processes, offering valuable insights for mitigating extreme weather events, such as localized torrential rainfall that contributes to significant precipitation accumulation \cite{houze2014cloud}.

As a result, \emph{the dynamics of these observables can be interpreted as the linear evolution of warm bubble-like patterns acting across various data fields}, characterized by \emph{emergence}, \emph{growth}, and \emph{decay} behaviors, which is particularly meaningful from a dynamical systems perspective within the Koopman operator framework.
Several natural questions then arise: 
(1) Which choice of observables is preferred?; (2) How do different observables affect the extracted Koopman modes?; (3) How can a low-dimensional representation of transient weather dynamics be constructed?
These questions are closely intertwined. In particular, the third question is directly related to \emph{model reduction}, which aims to construct a low-dimensional surrogate model by selecting an appropriate number of Koopman modes that effectively capture the underlying large-scale regional weather simulations generated by SCALE. In fact, developing an accurate reduced-order model naturally informs the first two issues (that is, (3) infers (1) and (2)); it indicates which observables (e.g., velocity, vorticity, humidity, etc.) capture the essential dynamics with lower performance loss, and it reveals how the choice of observables affects the extracted Koopman modes, as evaluated by the reconstruction results (e.g., spatial and temporal dynamics) of the warm bubble-like experiment.
Needless to say, KMD has the potential to provide novel, direct insights into these prominent transient modes (possibly associated with warm bubble-like patterns), enabling the forecast of the onset or dissipation of torrential rain.
 
\subsection{Novelty and Contributions}
More recently, several attempts have explored short-term weather dynamics by SCALE-RM simulation data for weather prediction and control. For example, \cite{ohtsuka2024convex} 
proposed a convex optimization-based approach for quantitative weather control within a numerical weather prediction (NWP) framework by perturbing the initial conditions (e.g., temperature or specific humidity) as control inputs, along with sensitivity analysis. Alternatively, the ensemble Kalman filter (EnKF) was applied in \cite{honda2025exploring} to investigate the intrinsic predictability limits of localized convective rainfall events.
A trial case study on humidity ratio data in weather simulations showed that accurate reconstruction is possible with a few Koopman modes \cite{zhang2025KoopmanMode}.
The present work focuses on extracting dominant transient modes in weather systems to enable model reduction and, to the best of our knowledge, represents one of the first applications of Koopman operator to SCALE weather analysis. Through the lens of DMD, this study aims to address large-scale short-term weather dynamics and capture the leading warm bubble-like patterns, providing a preparatory step toward future weather control and anomaly detection.

In the spirit of sparsity-promoting DMD (SPDMD) \cite{jovanovic2014sparsity}, which effectively examines coherent fluid modes using sparsity methods, we are thus motivated to extract a small number of dominant modes in transient weather dynamics.
In contrast to alternative methods such as SINDy 
\cite{brunton2016discovering} and the sparse discovery of informative Koopman-invariant subspaces \cite{pan2021sparsity}, SPDMD does not require identifying the governing equations or constructing a sparse regression model of the system dynamics. Instead, SPDMD induces sparsity regularization on the ``amplitudes" derived from DMD, \emph{transforming the selection of dominant transient modes into the identification of the main contributing amplitudes} via a sparse convex optimization problem. By tuning user-specified sparse weights, the method optimally penalizes the \emph{number of nonzero amplitudes} \cite{zhang2025sparsity,graff2020reduced,kou2017improved,tsolovikos2020estimation}, thereby retaining only the most relevant principal modes.
The major advantage of this approach is that it makes a trade-off between the quality of approximation (i.e., accuracy) and the number of modes (i.e., model complexity), even when the available data consist of short, finite-length snapshots.
Hence, concerning the captured dominant contributors, such as leading modes that may include warm bubble-like patterns, can potentially provide a data-based low-order model form that is both robust and interpretable, flexible for large-scale transient weather systems.
The main contributions of this paper are listed as follows:
\begin{enumerate}
\item[(i)] 
{\bf\emph{Transient weather dynamics and modes}}: 
Compared with prior climate studies focused on steady-state (long-term) behavior \cite{navarra2024variability,navarra2021estimation,zhang2025sparsity,hogg2020exponentially}, large-scale weather simulations derived as finite-length time series data are governed by transient dynamics. To examine the impact and role of transient behavior, we analyze short-term data snapshots of varying lengths corresponding to \emph{different stages of observable evolution} (initial vs.\ mid stages), enabling us to distinguish dominant Koopman modes associated with different phases of the dynamics.

 \item[(ii)] 
 {\bf\emph{Observables of interest}}: 
The transient weather system is modeled from a data-driven perspective, where different choices of observables correspond to different measured data fields. In particular, we focus on {scalar fields} such as the \emph{magnitude of velocity} and \emph{vorticity}, and demonstrate how the \emph{emergence} and \emph{growth} of warm bubble-like patterns are captured by the corresponding transient Koopman modes and eigenvalues.

\item[(iii)] 
{\bf\emph{Sparsity, accuracy, and model complexity}}: 
Using the SPDMD framework, we can make a trade-off between the \emph{accuracy}
(e.g., reconstruction error or performance loss) and \emph{model complexity}
(e.g., the number of retained modes or nonzero amplitudes)
by tuning the sparsity weight.
This provides a systematic and flexible mechanism for dominant mode selection,
offering advantages over DMD in reduced-order modeling of large-scale transient weather dynamics.
\end{enumerate}

\subsection{Structure}
The rest of this paper is structured as follows. 
Section~\ref{sec:literature-rew} provides a literature review of the operator-theoretic analysis of climate systems, differentiating from this study of transient weather dynamics. 
Section~\ref{sec:model-data} implements the SCALE weather simulation model and data sources. 
Section~\ref{sec:methods-SPDMD} presents the methodology, including the DMD and SPDMD. 
Section~\ref{sec:experiments} evaluates the effectiveness of these methods through three experimental results. 
Finally, Section~\ref{sec:conclusion-discussion} discusses the physical implications related to precipitation and concludes the paper.

\section{Related Work: Operator Methods for Long-Term Climate Systems}\label{sec:literature-rew}
Operator-theoretic methods for long-term Earth system in climate and atmospheric science have been intensively studied in the literature 
\cite{navarra2024variability,navarra2021estimation,hogg2020exponentially,sanchez2024advancing, wang2020extended,alexander2020operator,alexander2017kernel,lintner2023identification,froyland2021spectral,badza2024identifying,froyland2024revealing,giannakis2015spatiotemporal}, 
provided with the spatial and temporal mode decompositions of dynamical systems.
Wang et al. \cite{wang2020extended} proposed a non-parametric statistical approach called kernel analog forecasting (KAF), which combines the Kernel method and the Koopman operator to make long-term statistical ENSO prediction possible. 
This method uses industrial-era Indo-Pacific SST data for training and evaluates La Ni\~{n}a events. 
It also illuminates a critical property of the KAF method: under measure-preserving and ergodic dynamics, it consistently approximates the conditional expectation of observables evolved by the Koopman operator of the dynamical system, conditioned on the observed data at forecast initialization \cite{alexander2020operator}. 
The KAF method was also adopted in \cite{alexander2017kernel} to forecast the Madden-Julian oscillation (MJO) and boreal summer intraseasonal oscillation. 
In addition to KAF, Lintner et al. \cite{lintner2023identification} addressed MJO identification through the application of Koopman operator technique to derive a real-time multivariate MJO index, enhancing the characterization and predictability of the MJO. 
Notably, the spectral decomposition of the Koopman operator, with eigenvalues corresponding to mode periods, reveals a dominant intraseasonal mode with a distinct period. 
Shortly after, Froyland et al. \cite{froyland2021spectral} investigated how Koopman eigenfunctions can directly describe coherent climate phenomena, such as long-term monthly SST data for ENSO, offering greater dynamical consistency and enhanced physical interpretability.
The transfer operator, understood as the adjoint (or dual) of the Koopman operator in an appropriate function space \cite{lasota2013chaos}, is effective in identifying persistent structures in complex dynamical systems \cite{froyland2024revealing}—for instance, in turbulent flows where it has been used to detect atmospheric blocks \cite{badza2024identifying}. This motivates our investigation into structural emergence during the early stages of weather system's evolution and the disappearance of coherent sets \cite{atnip2024inflated}.
Meanwhile, Navarra et al.~\cite{navarra2021estimation} directly estimated the Koopman and transfer operators for the equatorial Pacific SST, using monthly mean SST data from the Ni\~{n}o~3 index in ERA5. Their approach employed the kernelized, extended DMD (EDMD)
method \cite{williams2015data} to derive a Gram matrix and benchmarked the results against EOF and LIM methodologies.
In particular, Navarra and his collaborators \cite{navarra2024variability} examined the \emph{continuous spectrum} (e.g., see~\cite{mezic2005spectral,colbrook2024rigorous}) of the Koopman operator to obtain the geographical distribution of dynamical modes, revealing SST variability through Koopman modes and emphasizing local oscillatory or decaying behavior within the climate system. The works \cite{sanchez2024advancing,lorenzo2025koopman} introduced Koopman ensemble forecasting for Pacific SST records and also taken shorter data into account.
Giannakis et al. \cite{giannakis2015spatiotemporal} analyzed the Koopman spatiotemporal modes of tropical convective variability across diurnal to seasonal timescales. 
Moreover, it is worth noting that many interesting and insightful studies have utilized the Koopman operator to investigate various climate phenomena, such as tropical intraseasonal oscillations \cite{alexander2020operator}, the variability of SST \cite{navarra2024variability,navarra2021estimation,sanchez2024advancing,wang2020extended}, 
and sea ice cover \cite{hogg2020exponentially}, among others that have yet to be mentioned.  
In \cite{zhang2025sparsity}, SPDMD was applied to a real-world SST dataset to identify the leading oscillation modes in the long-term climate dynamics. Most of the extracted eigenvalues lie on the unit circle, suggesting that the underlying climate dynamics are approximately measure-preserving \cite{froyland2021spectral} and predominantly in a steady state.

\begin{figure*}[!t]
    \centering
    \includegraphics[width=0.85\linewidth]{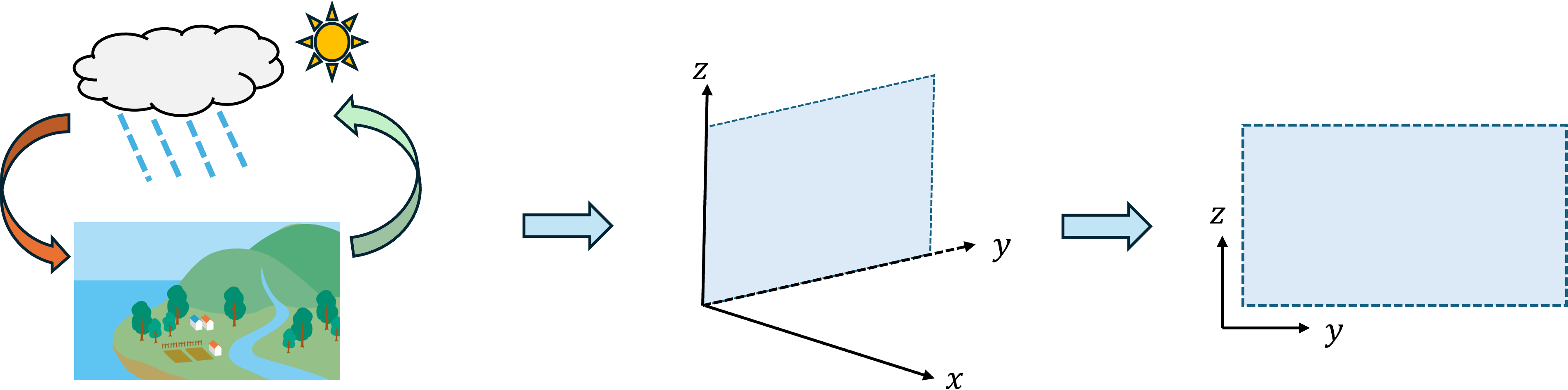}
    \caption{A sketch of SCALE weather simulation for warm-bubble-like experiment.}
    \label{fig:SCALE}
\end{figure*}

\section{Model and Data}\label{sec:model-data}
\subsection{Model Configuration}
In this study, we utilize weather simulation datasets generated by a regional model from the
SCALE {(or SCALE-RM)} ver 5.4.5 \cite{nishizawa2020scale}, a fundamental library for weather and climate modeling of the Earth
SCALE-RM solves the non-hydrostatic, three-dimensional, fully compressible governing equations. 
The model configuration largely follows that of \cite{ohtsuka2024convex} and \cite{nishizawa2020scale,honda2025exploring}. 
To simplify the analysis, this study does not include turbulence or radiation in weather simulations. 
More specifically, as shown in Fig.~\ref{fig:SCALE}, the model is configured to represent a two-dimensional $y$-$z$ plane with a horizontal grid resolution of $0.5$ km along the $y$ direction, spanning $40$ grid points, and $97$ vertical layers along the $z$ direction, extending up to an altitude of $20$ km. 
Therefore, the $y$-$z$ plane consists of a total of $40\times{97}=3880$ grid points, forming a single \emph{snapshot}.
In the warm bubble experiment performed using the SCALE weather simulation, the total simulation time is set to $T_s = 60$ minutes (equivalently, 3600 seconds), and the number of iterations or snapshots $N=121$, yielding a sampling interval (or time resolution) of $h:=T_s/(N-1)= 0.5$ minutes (or 30 seconds); see Table~\ref{tab:simulation_summary} for SCALE simulation setup details and Fig.~\ref{fig:SCALE} for a schematic overview.

\begin{table}[t!]
\centering
\caption{Setup of SCALE Weather Simulation}
\begin{tabular}{p{6cm} p{6.5cm}}  
\toprule
\textbf{Parameters} & \textbf{Units / Description} \\
\midrule
Simulation domain & Two-dimensional $y$-$z$ plane \\
Horizontal resolution $y$ & Up to $20$ km, $40$ points with $0.5$ km/grid \\
Vertical layers $z$ & Up to $20$ km, $97$ points with 0.21 km/grid \\
Total spatial grid points $p$& $p=n_y\times{n_z}=40 \times 97 = 3880$ grids\\
Total simulation time $T_{s}$& $3600$ seconds (or $60$ minutes, or $1$ hour) \\
Number of iterations $N$ & $121$ iterations/snapshots ($0.5$ min each)\\
Time resolution $h$& $30$ seconds per (or $0.5$ minutes each)\\
\bottomrule
\end{tabular}
\label{tab:simulation_summary}
\end{table}

\subsection{Data Source}\label{subsec:data-source}
 As depicted in Fig.~\ref{fig:SCALE}, the datasets used in this study include the \emph{horizontal velocity component} in the  $y$-direction, denoted by $\mathtt{data.V} := \mathtt{D}_{\mathtt{V}}$, and the \emph{vertical velocity component} in the $z$-direction, denoted by \(\mathtt{data.W} := \mathtt{D}_{\mathtt{W}}\).
For each snapshot $k = 0, \ldots, N-1$, these components define scalar fields
\begin{align*}
u_k^{y}(y_{m}, z_{\ell}), \quad u_k^{z}(y_{m}, z_{\ell}) \in \mathbb{R}, \quad \text{for} m=1,\ldots,n_y,\quad  \ell=1,\ldots,n_z,
\end{align*}
which correspond to the entries of $\mathtt{D}_{\mathtt{V}}$ and $\mathtt{D}_{\mathtt{W}}$, respectively. Each scalar field is defined on a two-dimensional spatial grid
$\{(y_{m}, z_{\ell}) \in \mathbb{R}^2 \mid m=1,\ldots,n_y; \; \ell=1,\ldots,n_z \}$,
where $n_y=40$ and $n_z=97$ denote the number of resolution points along the $y$ and $z$ directions with distance $20$~km, respectively.

Thus, the velocity components $u_k^{y}$ and $u_k^{z}$ are matrices of size $n_y \times n_z$ representing scalar fields over the spatial domain at snapshot $k$. For data-driven analysis such as dynamic mode decomposition, these matrices are flattened into vectors of length $p = n_y \times n_z = 3880$, producing the snapshot vectors
\begin{align*}
\mathbf{y}_k \in \mathbb{R}^p, \quad k=0, \ldots, N-1,
\end{align*}
with $N=121$ being the total number of snapshots in time. In particular, 
$\mathbf{y}_k$ are often referred to as the \emph{vector-valued} observables.
In this paper, we focus on the SCALE weather simulations acting on different observable spaces, each measured by the following data fields\footnote{The best track baseline and raw data, generated from SCALE weather simulations and including velocity and vorticity magnitudes, are freely 
\href{https://drive.google.com/drive/folders/1aIjP4sDFhO6NG296vyf7FubWPyPo0VtE?usp=drive_link}{accessible}; see the Appendix for data availability and the original snapshots (i.e., Figs.~\ref{fig:Ex-0-velocity-case-original-data} and~\ref{fig:Ex-1-Vorticity-case-original-data}).}:

\begin{itemize}
\item {\bf Velocity magnitude}: 
The SCALE weather data consists of a two-dimensional slice of a meteorological field, measured at $3880$ spatial grid points in the $y$-$z$ plane. 
The pointwise velocity magnitude at each grid for snapshot $k=0,1,\ldots,N-1$ is defined as
\begin{align*}
{\bm\Pi}_k(y_{m}, z_{\ell}) := \sqrt{ \big(u_k^{y}(y_{m}, z_{\ell})\big)^2 + \big(u_k^{z}(y_{m}, z_{\ell})\big)^2 },
\end{align*}
where $\mathbf{\Pi}_k(y_{m},z_{\ell})\in\mathbb{R}^{n_y\times{n}_z}$ defines the matrix in two-dimensional spatial grid. In order to collect the data snapshots for numerical analysis, this two-dimensional scalar field of velocity magnitude is further vectorized (flattened) into the column vector, defined as 
\begin{align*}
\bfy_k := \mathrm{vec}\bigl(\mathbf{\Pi}_k(y_m, z_{\ell})\bigr) \in \mathbb{R}^{p}, \quad \text{with } p = n_y \times n_z=3880, \quad k=0,1,\ldots,120.
\end{align*}
Fig.~\ref{fig:Ex-0-velocity-case-original-data} displays the partially selected original snapshots with respect to the velocity magnitude.
\item {\bf Vorticity magnitude}:
Another alternative observable data of the dynamic evolution in SCALE weather systems is the vorticity ${\bm\omega}$, defined as the curl of the velocity field, which is often used to quantify (point) vortex dynamics in the atmosphere and ocean \cite{pedlosky2013geophysical}.
For simplicity, we focus on the case of two-dimensional incompressible flow in the SCALE simulations, where the magnitude of vorticity reduces to a scalar field. 
Following the setting of velocity magnitude, 
for each snapshot $ k = 0, \ldots, N-1 $, the vorticity scalar field on the spatial grid is defined by
\begin{align*}
{\bm\Omega}_k(y_{m}, z_{\ell}):= \left|\frac{\partial u_k^{z}}{\partial y}(y_{m}, z_{\ell}) - \frac{\partial u_k^{y}}{\partial z}(y_{m}, z_j)\right|, \quad {m} = 1, \ldots, n_y;\,{\ell} = 1, \ldots, n_z,
\end{align*}
which quantifies the local rotational motion in the $y$-$z$ plane.
Finite differences can be used to compute these gradients.
The numerical gradients are stacked in the vector ${\bm \omega}_k$, which represents measurements $\bfy_k$ derived from the vorticity magnitude of the vortex dynamics. Formally, we define the vectorized snapshot of the vorticity magnitude in scalar field as:
\begin{align*}
    \bfy_k:={\bm\omega}_{k}=\mathrm{vec}({\bm\Omega}_k(y_{m}, z_{\ell}))\in\mathbb{R}^{p},
   \quad  p=n_y\times n_z =3880, \quad k=0,1,\ldots,120.
\end{align*}
Fig.~\ref{fig:Ex-1-Vorticity-case-original-data} presents partially selected original data of vorticity magnitude 
represented in the observed latent space over the $y$-$z$ plane from the SCALE weather simulation. 
Additional full snapshots and the movie are available in Subsection~\ref{subsec:data-source}. These snapshots reveal the evolution of warm bubble-like patterns rising upward, 
implying the onset or dissipation of heavy rainfall.
 \end{itemize}

 \begin{figure*}[!t]
    \centering
     \includegraphics[width=\linewidth]{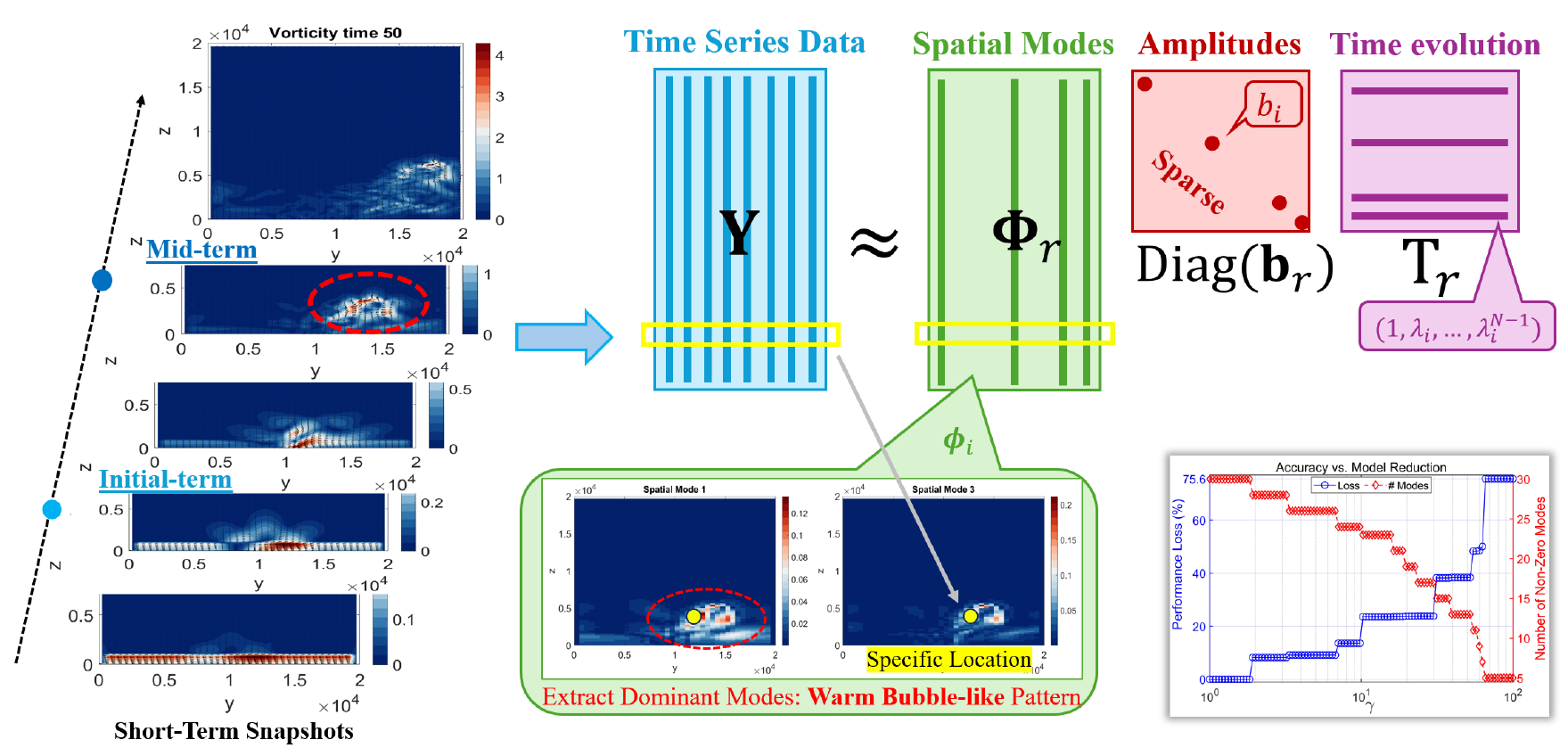}  
    \caption{SPDMD roadmap: extracting dominant modes with warm bubble-like patterns.}
    \label{fig:roadmap-SPDMD}
\end{figure*}

%
\section{Methodology}\label{sec:methods-SPDMD}
In this study, we restrict our attention to the ``{point spectrum}" of the Koopman operator underlying the Koopman mode decomposition (KMD), focusing on coherent structures (e.g., warm bubble-like pattern) in transient weather dynamics measured from finite-length snapshots.
For a rigorous treatment of continuous Koopman spectra, including cases exhibiting mixing, chaotic, or stochastic behavior, the interested reader is referred to \cite{mezic2005spectral,colbrook2024rigorous}.
In what follows, we will recall the DMD methods (see, e.g., in~\cite{rowley2009spectral,schmid2010dynamic,kutz2016dynamic}) to estimate the eigenvalues of the Koopman operator and the associated Koopman modes.
In particular, this work emphasizes model reduction using SPDMD \cite{jovanovic2014sparsity}, which induces sparse amplitudes in DMD.
Fig.~\ref{fig:roadmap-SPDMD} illustrates the roadmap of the SPDMD method for extracting dominant modes from short-term SCALE weather data. 
\subsection{Dynamic Mode Decomposition}
First, we review the DMD architecture \cite[Chapter~1]{kutz2016dynamic}, a widely used data-driven method for approximating Koopman eigenvalues and Koopman modes via a finite-dimensional eigen-decomposition constructed from snapshots of system trajectories.
We record finite-length time-series data directly from SCALE weather simulations, measured by
\begin{align*}
    \left\{\bfy_{0},\bfy_{1},\ldots,\bfy_{N}\right\},
\end{align*}
where $\bfy_{k}\in\mathbb{R}^{p}$ is the snapshot of the SCALE simulation at discrete time $k$, $p$ is the number of measurement locations or sensors,  
and $N+1$ is the number of accessible snapshots with $N\ll{p}$. 
Subsequently, we collect the following measured data matrices 
\begin{equation}
\begin{aligned}
    \mathbf{Y}:=
    \begin{bmatrix}
        \bfy_{0} & \bfy_{1} & \ldots & \bfy_{N-1}
    \end{bmatrix}
    \quad \text{and} \quad 
    \mathbf{Y}'=\begin{bmatrix}
        \bfy_{1} & \bfy_{2} & \ldots & \bfy_{N}
    \end{bmatrix},
\end{aligned}
\label{eq:separate-data-Y}
\end{equation}
where $\bfY,\bfY'\in\mathbb{R}^{p\times{N}}$, $\bfY'$ represents a time shifted data matrix of $\bfY$.
For this, we find 
discrete-time linear time-invariant dynamics, expressed by 
\begin{align}
\bfy_{k+1}\approx
\bfA\bfy_{k},\quad 
\Longrightarrow\quad 
\bfY'\approx
\bfA\bfY,
\label{eq:DT_LTI_DMD}
\end{align}
where $\bfA$ is a constant matrix that approximates the Koopman operator in a finite-dimensional form\footnote{Throughout this paper, we consider only the point spectrum (i.e., discrete eigenvalues) of the finite-dimensional matrix $\bfA$ obtained via DMD, see \eqref{eq:DMD-1}. While the underlying Koopman operator is infinite-dimensional and may contain continuous spectrum \cite{mezic2005spectral}, this aspect is beyond the scope of the present work and the (SP)DMD frameworks.} (e.g., see \cite{mauroy2020koopman,tu2014dynamic,brunton2022data} in detail).
DMD is a numerical method of KMD that minimizes the least-square errors in the linear model \eqref{eq:DT_LTI_DMD}:
\begin{equation}
\min_{\bfA\in\mathbb{R}^{p\times p}}\|\bfY'-\bfA\bfY\|_{\rm F},
\label{eqn:LSprob}
\end{equation}
where $\|\cdot\|_{\rm F}$ is the Frobenius norm of matrices. 
This yields the proper solution $\bfA =\bfY'(\bfY)^{\dag}$, where $(\cdot)^{\dag}$ is the Moore-Penrose pseudoinverse.
In practice, such an exact solution is often not directly tractable, especially in high-dimensional weather systems. Instead, we prefer to compute a reduced-order estimation $\tilde{\bfA}\in\mathbb{C}^{r\times{r}}$ of the full-order operator $\bfA\in\mathbb{R}^{p\times{p}}$. To achieve this, we apply the singular value decomposition (SVD) to the original data matrix $\bfY\in\mathbb{R}^{p\times{N}}$ (i.e., $\bfY=\bfU{\bmSigma}{\bfV}^{\ast}$), which further leads to a rank-$r$ truncated approximation:
\begin{align*}
   \bfY\approx \bfU_r\bmSigma_r{\bfV}_r^{\ast}
   \quad  \text{with}\quad \bfU_r^{\ast}\bfU_r=\bfI, \quad \bfV_r^{\ast}\bfV_r=\bfI,
\end{align*}
where $r\leq\min\{p,N\}$. 
Here, $\bfU_r\in\mathbb{R}^{p\times{r}}$ and  $\bfV_r\in\mathbb{R}^{N\times{r}}$ contain the leading $r$ left and right singular vectors of $\bfY$, respectively, 
and $\bmSigma_r\in\mathbb{R}^{r\times{r}}$ is the diagonal matrix with the top $r$ singular values. 
Notice that the columns of $\bfU_r$ are the orthonormal eigenvectors of $\bfY\bfY^{\ast}$, or equivalently, the proper orthogonal decomposition (POD) modes \cite{holmes1996turbulence,schmid2010dynamic}.
Using this decomposition (i.e., $\bfA=\bfY'(\bfY)^{\dag}\approx\bfY^{'}\bfV_{r}\bmSigma_{r}^{-1}\bfU_r^{\ast}$), we construct a reduced operator via projection:
\begin{align}
\tilde{\bfA}=\bfU_r^{\ast}\bfA\bfU_r\in\mathbb{R}^{r\times{r}},
    \quad \Longrightarrow\quad
    \tilde{\bfA}\approx \bfU_{r}^{\ast}\bfY^{'}\bfV_{r}\bmSigma_{r}^{-1},
    \label{eq:decomposition-tilde_A}
\end{align}
which captures the dominant low-rank dynamics $\tilde{\bfy}_{k+1}\approx\tilde{\bfA}\tilde{\bfy}_{k}$ of the original system \eqref{eq:DT_LTI_DMD}.

Suppose that $\tilde{\bfA}$ has a finite (countable) set of linearly independent (right) eigenvectors $\bfw_{j}$ and the associated eigenvalues $\lambda_{j}$ such that $\tilde{\bfA}\bfw_{j}=\lambda_{j}\bfw_{j}$,
then the compact form is given as follows 
\begin{align*}
    \tilde{\bfA}\bfW={\bmLambda}\bfW,\quad \text{with}\quad 
    \bfW=[\bfw_{1}~\cdots~ \bfw_{r}]
    \in\mathbb{C}^{r\times{r}},\quad
     \bmLambda=\mathrm{diag}(\lambda_{1}\ldots\lambda_{r})\in\mathbb{C}^{r\times{r}}.
\end{align*}
By employing the eigen-decomposition of $\tilde{\bfA}$ as $\bfW\bmLambda{\bfW}^{-1}$, we obtain
\begin{align}
     \tilde{\bfy}_{k+1}\approx{\bfW}{\bmLambda}\bfW^{-1}\tilde{\bfy}_{k}
    ={\bfW}{\bmLambda}^{k}{\bfW}^{-1}\tilde{\bfy}_{0},
    \label{eq:eigen-decomposition-linear-tilde_A}
\end{align}
where $\tilde{\bfy}_0$ is the initial condition. Let $\bfz_j$ 
be the $j$-th (left) eigenvector of $\tilde{\bfA}$ satisfying $\bfz^{\ast}_j\tilde{\bfA}=\lambda_j{\bfz}_j^{\ast}$ (i.e., the eigenvector of $\tilde{\bfA}^{\ast}$), or equivalently in the compact form $\bfZ^{\ast}\tilde{\bfA}=\bmLambda{\bfZ}^{\ast}$ with $\bfZ^{\ast}=[\bfz_1^{\ast}\ldots{\bfz}^{\ast}_r]\in\mathbb{C}^{r\times{r}}$. This implies the bi-orthogonality condition between the left and right eigenvectors $\bfZ^{\ast}\bfW=\bfI$, i.e.,
\begin{align*}
    \bfz^{\ast}_{i}\bfw_{j}=\delta_{ij}=
    \begin{cases}
        1,\quad i=j,\\
        0,\quad i\neq{j},
    \end{cases}
\end{align*}
where $\delta_{ij}$ is the Kronecker delta. According to \eqref{eq:eigen-decomposition-linear-tilde_A}, we have the following solution form
\begin{align*}
\tilde{\bfy}_{k}= \bfW\bmLambda^{k}\bfZ^{\ast}\tilde{\bfy}_{0}=\sum_{j=1}^{r}\bfw_{j}\lambda_{j}^{k}\bfz^{\ast}_j\tilde{\bfy}_{0}=\sum_{j=1}^{r}\bfw_{j}\lambda_{j}^{k}b_j,
\end{align*}
where $b_j:=\bfz^{\ast}_j\tilde{\bfy}_{0}$ represents the $j$-th modal contribution of the initial condition $\tilde{\bfy}_{0}$.
Furthermore, the matrix of POD modes $\bfU_r$ can be used to map $\tilde{\bfy}_{k}$ into a higher-dimensional space $\mathbb{C}^{p}$ by $\bfy_{k}\approx{\bfU}_{r}\tilde{\bfy}_{k}$. 
We can approximate raw data snapshots using a linear combination of the DMD modes 
${\bm\phi}_{j}=\bfU_{r}\bfw_j$\footnote{Under the point spectrum assumption, the Koopman operator admits $\mathcal{K}^{k}{\bm\varphi}_j=\lambda_j{\bm\varphi}_j$ \cite{mezic2005spectral,mauroy2020koopman}, and thus KMD can be expressed as a finite truncation of the spectral expansion:
$\bfy_k={\bm{g}}(\bfx_k)\approx\sum_{j=1}^{N_r}\lambda_j^{k}{\bm\varphi}_j(\bfx_0)\bmV_j$,
where $\lambda_j$ are the Koopman eigenvalues, 
${\bm\varphi}_j$ denote the Koopman eigenfunctions, 
and $\bmV_j:=\langle{\bm\varphi}_j,{\bm{g}}\rangle$ are the Koopman modes associated with the vector-valued observable ${\bm{g}}\in\mathbb{R}^{p}$. As discussed in \cite[Section~4]{tu2014dynamic}, KMD is directly connected to DMD.}:
\begin{align}
 \bfy_{k}\approx{\bfU}_r\tilde{\bfy}_{k}
 =\sum_{j=1}^{r}{\bm\phi}_{j}\lambda_{j}^{k}b_{j}, 
 \quad \bfy_{0}\approx \sum_{j=1}^{r}{\bm\phi}_{j}b_{j}, 
\label{eq:DMD-1}
\end{align}
where $b_j$ is interpreted as the \emph{amplitude} of DMD modes \cite{schmid2010dynamic}. 
From \eqref{eq:DMD-1}, we derive a compact form:
\begin{align}
    \bfy_{k+1}\approx{\bm\Phi}_{r}{\bmLambda}{\bm\Phi}_{r}^{-1}\bfy_{k}
    ={\bm\Phi}_{r}{\bmLambda}^{k}{\bm\Phi}_{r}^{-1}\bfy_{0}
    :={\bm\Phi}_{r}{\bmLambda}^{k}\bfb_{r}.
    \label{eq:eigen-decomposition-DMD}
\end{align}
Here $\bfb_r:= [b_1, \ldots, b_r]^\top = {\bm\Phi}_{r}^{-1} \bfy_0 \in \mathbb{C}^{r \times 1}$ is the vector of DMD amplitudes, and 
${\bm\Phi}_{r}:= [{\bm \phi}_1 \ldots {\bm\phi}_r] \in \mathbb{C}^{p \times r}$ indicates the matrix of DMD modes.
As stated in \cite{schmid2010dynamic,tu2014dynamic}, the eigenvalues $\lambda_j$ are regarded as the estimated Koopman eigenvalues, and ${\bm\phi}_j$ are viewed as the estimated Koopman modes for $j=1,\ldots,r$. 
Finally, by going back to \eqref{eq:eigen-decomposition-DMD} while combing it with the snapshot data matrix $\bfY$, we see
\begin{align}
\bfY:=\begin{bmatrix}
       |     & |     & \cdots & |\\
      \bfy_{0} & \bfy_{1} & \cdots &\bfy_{N-1}  \\
       |     & |     & \cdots & |
     \end{bmatrix}
   \approx
   \underbrace{\begin{bmatrix}
      |    & | & \cdots & |\\
      {\bm \phi}_{1} & {\bm \phi}_{2} & \cdots & {\bm \phi}_{r}  \\
        |    & |&  \cdots & |
    \end{bmatrix}}_{{\bm\Phi}_r} 
    \underbrace{\begin{bmatrix}
        b_{1}  & &\\
        & \ddots&\\
        & & b_{r}
    \end{bmatrix}}_{\mathrm{diag}(\bfb_r)}
    \underbrace{\begin{bmatrix}
        1 & \lambda_{1} & \ldots & \lambda_{1}^{N-1}\\
         \vdots & \vdots & \ddots &\vdots\\
          1 & \lambda_{r} & \ldots & \lambda_{r}^{N-1}\\
    \end{bmatrix}}_{\bfT_r},
    \label{eq:dynamic-mode-decomposition-1}
\end{align}
where $\mathrm{diag}(\bfb_r)\in\mathbb{C}^{r\times{r}}$ is the diagonal matrix of the amplitudes,
and $\mathbf{\Phi}_{r}\in \mathbb{C}^{p \times r}$ represents the DMD spatial modes, computed by using either ${\bm\Phi}_{r}=\bfU_{r}\bfW$, or alternatively, ${\bm\Phi}_{r}=\bfY'\bfV_{r}\bmSigma_{r}^{-1}\bfW$, and $\bfT_r\in\mathbb{C}^{r\times{N}}$ is the Vandermonde matrix built from the eigenvalues describing the temporal modes. 
The \emph{unknown} amplitude vector $\bfb_{r}$ is obtained by solving the following optimization problem\footnote{By using economy-size (or $r$-truncated) SVD of $\bfY\approx\bfU_r\bmSigma_r{\bfV}_r^{\ast}$ and observing that ${\bm\Phi}_{r}=\bfU_rW$, the optimal solution to the problem \eqref{eq:least-square-prob} equals to $\bfb_{r}\in\arg\min\|\bmSigma_r{\bfV}_r^{\ast}-\bfW\mathrm{diag}(\bfb_r)\bfT_{r}\|_{\text{F}}^{2}$. It can be recast as a convex quadratic program:
\begin{align}
    J(\bfb_r)=\bfb_r^{\ast}\mathbf{P}\bfb_r-\bfb_{r}^{\ast}\mathbf{d}-\mathbf{d}^{\ast}\bfb_{r} + \eta,
    \label{eq:QP-closed-form-solution}
\end{align} 
with a straightforward and closed-form solution 
$\bfb_r=\mathbf{P}^{-1}\mathbf{d}$, where $\mathbf{P}=(\bfW^{\ast}\bfW)\circ(\overline{{\bfT}_{r}{\bfT}_{r}^{\ast}})$,
$\mathbf{d}=\overline{\mathrm{vdiag}(\bfT_{r}\bfV_r\bmSigma_r^{\ast}\bfW)}$, and $\eta=\mathrm{trace}(\bmSigma_r^{\ast}\bmSigma_r)$, see \cite[Appendix~A]{jovanovic2014sparsity}. The symbol $\circ$ implies element-wise multiplication,
$\overline{M}$ is the conjugate of $M$, $M^\ast$ is the conjugate transpose of $M$, and $\mathrm{vdiag}(M)$ is the vector consisting of the diagonal elements of the square matrix $M$.}: 
\begin{align}
   \min_{\bfb_{r}\in\mathbb{C}^{r}}
   \|\bfY-{{\bm\Phi}_r\mathrm{diag}(\bfb_r)\bfT_r}
   \|_{\text{F}}^{2}.
    \label{eq:least-square-prob}
\end{align}

A superposition of all DMD modes, appropriately weighted by their amplitude and advanced in time according to their temporal growth/decay rate, provides an optimal approximation of the original data sequence $\bfy_k$ and recovers a normalized scalar field in the $y$-$z$ plane by assuming that the norm $\bm\phi_j$ is scaled at unity.
In this context, $\bfy_k$ are the vectors obtained by flattening the scalar-field data on the $y$-$z$ plane, and each mode ${\bm \phi}_j$ represents the spatial pattern of single-frequency oscillation, which is characterized by the temporal evolution $\lambda^k_jb_j$ in discrete-time (or, in the continuous-time case with sampling interval $\Delta{t=1}$, $\mu_j=\log(\lambda_j)/\Delta{t}$), embedded in the field data. 
The so-called normal evolution of single-mode in real-valued time series are then defined as  
\begin{align}
   a_{k}^{\text{mode}~j}=\mathrm{Re}(\lambda^k_jb_j).
    \label{eq:normal-evolution}
\end{align}
If the absolute value (also known as modulus or magnitude) $|\lambda_j|$ is equal to unity (except for $\lambda_j\neq{1}$), 
then the normal evolution
$a_{k}^{\text{mode}~j}$ exhibits a steady state, sustained oscillation. Otherwise, if the absolute value $|\lambda_j|$ greater (or small) than unity suggests that $a_{k}^{\text{mode}~j}$ will grow (or decay) over time, which indicates the \emph{transient} state with growth (or decay) mode.

Another aspect of interest in the system's evolution is its behavior at a specific spatial location, denoted by ${\bm\phi}_j^{\text{locat}~(y_{m},z_{\ell})}$, which lies on the $40 \times 97$ grid in the $y$–$z$ plane. This location, $(y_{m},z_{\ell})$, corresponds to a ``{highlighted index} $\mathrm{i}$" in the flattened state vector $\bfy_k \in \mathbb{R}^{3880 \times 1}$, representing the $\mathrm{i}$-th component of $\bfy_k$, denoted as $\bfy_{k,\mathrm{i}}$. Here, we examine the time evolutions of the sum of real values of the selected $r$ modes, describing their {superposition} at this specific spatial location, characterized by:
\begin{align} 
\bfy_{k,\rm{i}} \approx \sum_{j=1}^{r} \mathrm{Re}\!\left(\big[{\bm\phi}_{j}\big]^{\rm{i}} \lambda_{j}^{k} b_j\right), \quad \big[{\bm\phi}_{j}\big]^{\rm{i}} := {\bm\phi}_{j}^{{~\text{locat}~(y_{m},z_{\ell})}},
\label{eq:specific-loaction-superposition} 
\end{align} 
for ${m}=1,\cdots,n_{y},{\ell}=1,\ldots,n_z$.
Here $\mathrm{i}\in\{1,\ldots,p\}$ denotes a fixed flattened spatial index corresponding to the location
$(y_{m},z_{\ell})$ in the $y$-$z$ plane.
It is clear that the real-valued superposition's evolution \eqref{eq:specific-loaction-superposition} is based on the normal evolution \eqref{eq:normal-evolution}, but differs slightly from it. Specifically, it represents the dynamics of a low-dimensional system constructed from the sum of selected $r$ modes, enabling comparisons at the same locations as the original weather data field.

\subsection{Sparsity-promoting dynamic mode decomposition}
Above, we derived the decomposition into the $r$ spatial modes ${\bm\phi}_j$ with single frequencies characterized by $\lambda_j$. 
Here, to minimize the number of modes for low-order approximation of the data, we introduce the least-square error problem \eqref{eq:least-square-prob}
with $\ell_{1}$ regularization (a.k.a., sparse, or Lasso constraint) as
\begin{align}
   \min_{{\bfb_r}\in\mathbb{C}^{r}}
   \|\bfY-{\bf\Phi}_r\mathrm{diag}(\bfb_r)\bfT_{r}
   \|_{\text{F}}^{2}
   \quad \text{s.t.}\quad \|{\bfb_r}\|_{1}\leq{s},
    \label{eq:least-square-prob-L1-regularization}
\end{align}
where $\|\bfb_r\|_{1}:=\sum_{i=1}^{r}|b_{i}|$ is the $\ell_{1}$ norm, and $s\in\mathbb{R}$ is the tuning parameter that can be adjusted to achieve a desired sparsity level. 
As mentioned in \cite{kou2017improved,tsolovikos2020estimation}, the constrained convex program \eqref{eq:least-square-prob-L1-regularization} can be recast into the regularized problem \eqref{eq:spDMD}.

Jovanovi\'{c} \emph{et al.}~\cite{jovanovic2014sparsity} proposed the \emph{sparsity-promoting} dynamic mode decomposition (SPDMD) that minimizes the approximation error and additional $\ell_{1}$-norm penalty as
\begin{align}
    {J}_{\gamma}(\bfb_r)=\|\bfY-\mathbf{\Phi}_r\mathrm{diag}(\bfb_r)
    \bfT_{r}
    \|_{\mathrm{F}}^{2}+\gamma\|\bfb_r\|_{1},
    \label{eq:spDMD}
\end{align}
where $\gamma\geq{0}$ is the weight to tune the sparsity level that makes a trade-off between the accuracy and the number of reduced or selected modes. 
SPDMD identifies a low-dimensional linear system representation that captures the most dominant dynamic modes while discarding features that contribute weakly to the data sequence. This method extends the standard DMD algorithm to derive \eqref{eq:dynamic-mode-decomposition-1} solely.
As discussed in Remark~\ref{remark:role-of-sparsity}, the main feature of sparsity is the tendency to return \emph{sparse solutions for the amplitudes}, i.e., solutions with many zero components $b_j$. 

Let $(\lambda_{\varsigma_j},\bm{\phi}_{\varsigma_j})$, $j=1,\ldots,s$, denote the selected pairs of Koopman eigenvalue and mode, obtained by ordering the \emph{magnitude of non-zero amplitudes} $|b_{\varsigma_j}|$ via sparse optimization \eqref{eq:spDMD} such that 
\begin{align}
    |b_{\varsigma_{1}}|\geq|b_{\varsigma_{2}}|\geq\cdots\geq|{b}_{\varsigma_{s}}|> 0,\qquad |b_{\varsigma_{s+1}}|=\cdots=|b_{\varsigma_{r}}|=0,
    \label{eqn:order-amplitudes-b}
\end{align}
where $s\ll{r}$ and the rest of $r-s$ amplitudes are induced to zero by the sparsity-promoting regularization imposed on the amplitude vector $\mathbf{b}_r$ (i.e., taking $\ell_1$ norm $\|{\bf{b}}_r\|_{1}\leq{s}$ in \eqref{eq:least-square-prob-L1-regularization}).
Notice that the parameter $\gamma$ in \eqref{eq:spDMD} serves a similar role to $s$ in problem \eqref{eq:least-square-prob-L1-regularization}. In other words, there exists a re-ranking strategy $\varsigma(\gamma):\{1,\ldots,r\}\to\{\varsigma_{1},\ldots,\varsigma_{s}\}$ that induces the ordering given in \eqref{eqn:order-amplitudes-b}.
This method is to recover the good reconstruction using as few modes $s$ as possible. This is achieved by the SPDMD-based re-superposition, where the reconstructed time-series data snapshot $\widehat{\bfy}_{k,\rm{i}}$ at time $k$ is expressed by the real part 
\begin{align}
{\bfy}_{k,\rm{i}}:=\widehat{\bfy}_{k,\rm{i}} \approx \sum_{j=1}^{\varsigma_s}\mathrm{Re}\!\left([\bm{\phi}_{\varsigma_j}]^{\rm{i}}\lambda_{\varsigma_j}^kb_{\varsigma_j}\right),
\label{eqn:reconst-SPDMD-amplitudes-order}
\end{align}
where the index $\rm{i}$ denotes the spatial location within the domain, with $p=n_y{n}_z$. 
In view of this, the residual error can be evaluated by comparing the original time-series data $\bfy_{k,\rm{i}}$ defined in \eqref{eq:separate-data-Y}, with the SPDMD-based reconstruction given in \eqref{eqn:reconst-SPDMD-amplitudes-order}. 
The choice of dominant modes was addressed in \cite{jovanovic2014sparsity}, leading to the SPDMD.
For more details about SPDMD, we refer to see Algorithm~\ref{algo:spdmd} in Appendix.

\begin{remark}[The use of $\ell_{1}$ regularization]\label{remark:role-of-sparsity}
 Given a weight $\gamma\geq{0}$ that varies 
 continuously over $\mathbb{R}$, the solution ${\bfb_r^{\star}}(\gamma)$ forms a piecewise linear path as a function of $s$. 
 This is a consequence of the fact that ${\bfb_r^{\star}}(\gamma)$ is a unique minimizer of convex quadratic function $\|\bfY-\mathbf{\Phi}_r\mathrm{diag}(\bfb_r)\bfT_{r}
 \|_{\text{F}}^{2}$ over a closed domain with polyhedral constraint set $\{\|{\bfb_r}\|_{1}\leq{s}\}\cap\mathbb{C}^{r}$.
As $s$ increases, the feasible set expands, and the active set of the $\ell_{1}$ norm constraint changes at specific thresholds, resulting in a piecewise linear trajectory for $\bfb_r^{\star}(\gamma)$. 
In practice, this path is observed by solving the optimization problem for a range of $s$ values, typically sampled from the interval $[s_{\min},s_{\max}]$ with a chosen grid point. 
Each grid $s$ generates a related sparse solution $\bfb_r^{\star}(\gamma)$, resulting in a discrete approximation of the continuous path. 
The path of $\bfb_r^{\star}(\gamma)$ with $\gamma\in[\gamma_{\min},\gamma_{\max}]$ in problem \eqref{eq:spDMD} is the same to the above statement.
\end{remark}

\begin{remark}[Spectral distribution in SPDMD]\label{remark:SPDMD-spectral-distribution}
The SPDMD algorithm has no intrinsic preference for eigenvalues in any specific region of the complex plane (e.g., inside/outside the unit circle or left/right half-plane), since SPDMD is imposed on the modal amplitudes, not on the eigenvalues. Specifically, \emph{SPDMD extracts a subset of the eigenvalues obtained from DMD} by applying an $\ell_1$ penalty to the amplitude vector in \eqref{eq:least-square-prob-L1-regularization}, yielding 
$\{\lambda_{\varsigma_j}\}_{j=1}^{s} \subseteq \{\lambda_j\}_{j=1}^{r}$ with $s \le r$, as shown in \eqref{eqn:order-amplitudes-b}.
Hence, the resulting spectral distribution is entirely data-driven and primarily determined by amplitude magnitudes.
\end{remark}

\begin{figure*}[thb!] 
\centering 
\includegraphics[width=0.4\linewidth]{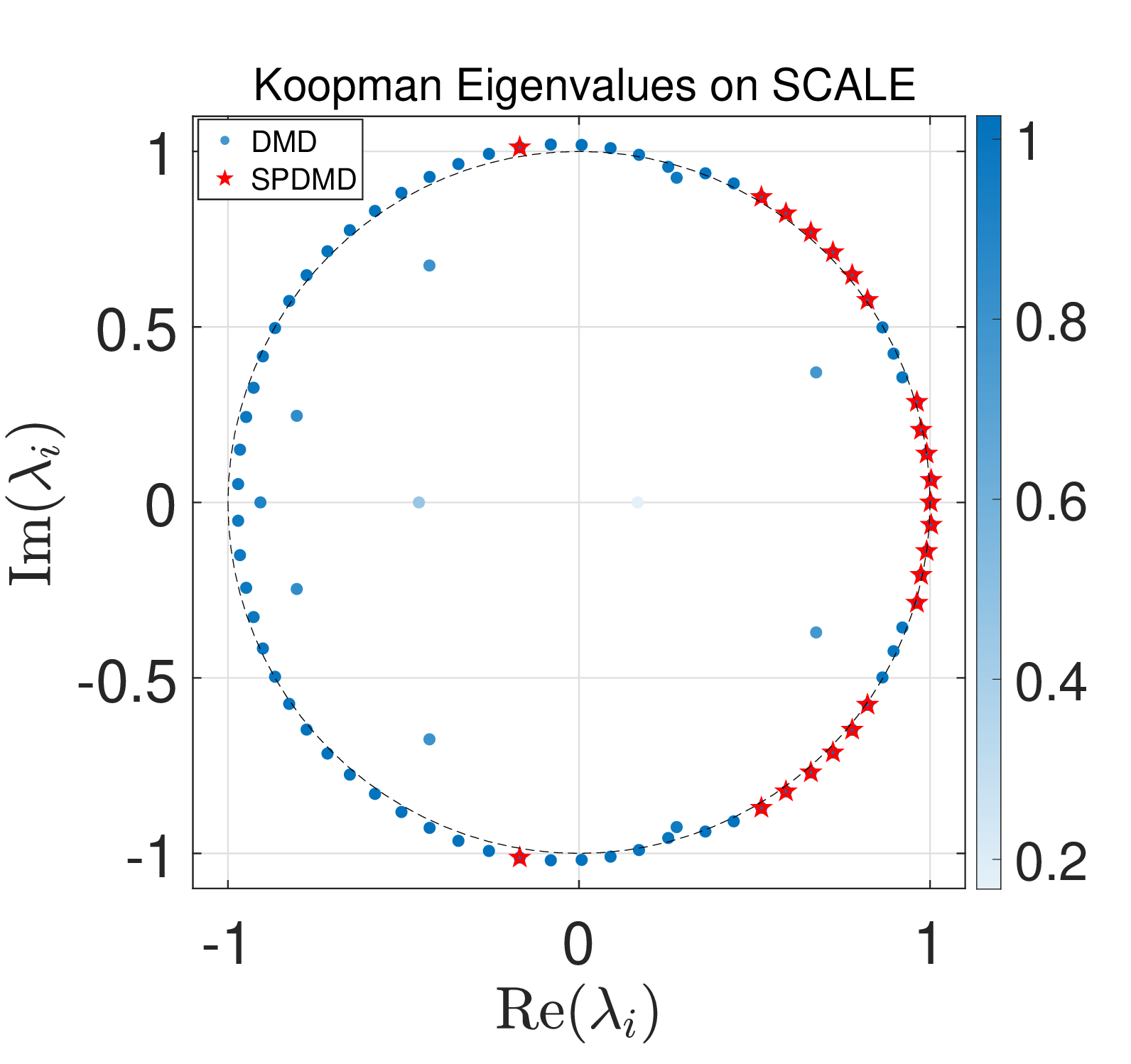}\quad
\includegraphics[width=0.4\linewidth]{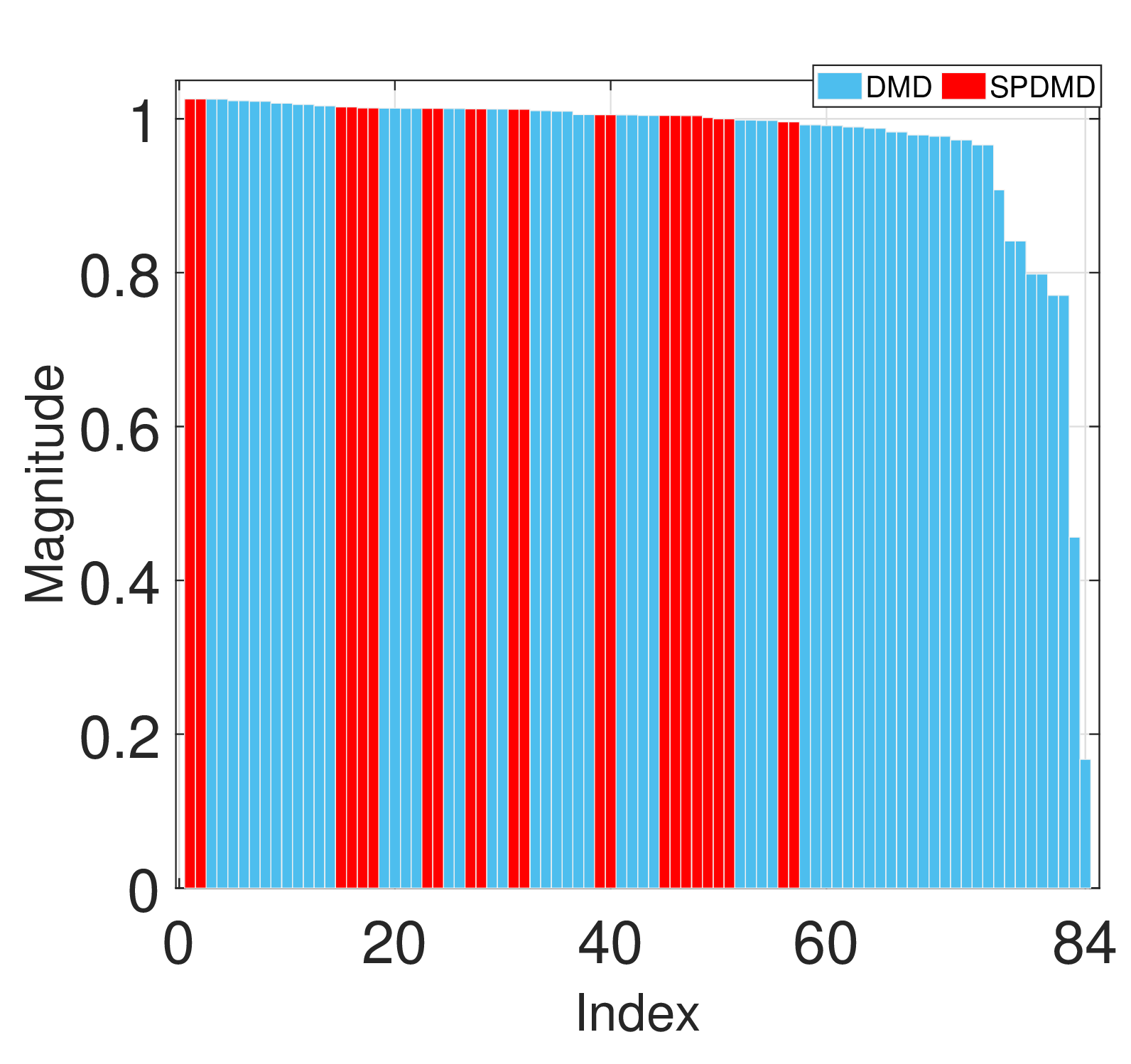}
\caption{Distribution and absolute values of Koopman eigenvalues estimated with the DMD and SPDMD methods for the data on velocity magnitude. 
The red stars stand for the SPDMD-based estimation, and the cyan points for the DMD-based estimation. 
In the left figure, plotted on the complex plane, the color bar shows the absolute values of the estimated eigenvalues from the DMD method.
In the right figure, the horizontal axis represents the indices of the eigenvalues estimated by the DMD method (shown as cyan bars), while the red bars indicate those selected by SPDMD based on amplitude magnitudes. The corresponding Table~\ref{tabel:scalar-mode-norm-periodic-84} provides additional details on the eigenvalues.}
\label{fig:Ex-0-scalar-eigenvalues}
\end{figure*} 

\section{Experiments and Results}\label{sec:experiments}
In this section, we
utilize the DMD and SPDMD for SCALE simulations
to extract the dominant behavior of transient weather dynamics, characterized by warm, bubble-like patterns, possibly leading to a low-dimensional linear representation of the original high-dimensional system.

\subsection{Velocity Magnitude}
We first show the scenario of the velocity magnitude in the scalar field. 
We conduct a short-term weather experiment using $N=85$ snapshots (corresponding to $T_s=42.5$ minutes with a time resolution $h=0.5$ minutes in SCALE weather simulations, i.e., $N=T_s/h$) to develop and demonstrate our Koopman analysis.
The eigenvalues estimated with the two methods (i.e., DMD and SPDMD) are presented in Fig.~\ref{fig:Ex-0-scalar-eigenvalues} and Table~\ref{tabel:scalar-mode-norm-periodic-84}. 
The left part of Fig.~\ref{fig:Ex-0-scalar-eigenvalues} shows the distributions and absolute values of the estimated eigenvalues on the complex plane, where the ``red stars" indicate the subset of eigenvalues selected by SPDMD from the set of eigenvalues computed by DMD (cyan points). This selection is performed by sparsity promoting on the modal amplitudes, which retains modes and the associated eigenvalues with the largest amplitude magnitudes; see \eqref{eqn:order-amplitudes-b} and Remark~\ref{remark:SPDMD-spectral-distribution} for details. The right part of Fig.~\ref{fig:Ex-0-scalar-eigenvalues} shows the absolute values ranked by their magnitudes. We can see here that most of the eigenvalues have their absolute values different from unity. 
Although this is partly due to numerical issues, it is mainly because we address transient data in the SCALE simulation, which differs from the long-term SST data as in \cite{navarra2024variability,navarra2021estimation}. 
In particular, as in Table~\ref{tabel:scalar-mode-norm-periodic-84} for SPDMD-based selection, except for Mode's label~1, the selected eigenvalues imply transient states, suggesting that the SCALE simulation here requires a combination of \emph{non-stationary} modes to capture the low-order systems.

\begin{table}[!b] 
\centering 
\begin{tabular}{@{\extracolsep{5pt}} c c c c c} 
\\[-1.8ex]\hline 
\hline \\[-1.8ex]
 Mode's label & DMD Amp. & Absolute value 
 & Eigenvalue  &  Period ($0.5$~min)\\ 
   $i$  & $|b_{i}|$ & $|\lambda_{i}|$ & $\lambda_{i}$ & ${2\pi}/{\mathrm{Im}}(\log\lambda_{i})$\\
\hline \\[-1.8ex] 
  \rowcolor{lightgray}
1 (1)  & 295.91 &  1.00 &  1.001 + 0.000i   & Inf \\
3  (2) & 25.72  & 1.01    & 1.003 + 0.064i  &  98.80\\
5 (5)  & 20.56   &  1.00   &  0.990 -- 0.139i   &  45.05\\
  \rowcolor{lightgray}
7 (6)   & 16.82   &  0.99  &  0.974 + 0.207i    &  30.5\\
9 (8)   & 9.98   &  1.01   &  0.963 + 0.286i   &  21.74\\
11 (13)  & 1.49   & 1.01    &  0.778 -- 0.648i   &  9.05\\
13 (17)  & 1.27   &  1.02   &  0.723 -- 0.712i     &  8.08\\
15 (20)  & 0.75   & 1.01    &  0.661 + 0.769i     & 7.30 \\
17 (19)   & 0.48   &  1.01   &  0.882 -- 0.577i   &  21.74\\
19 (23)  & 0.33   & 1.01    &  0.589 -- 0.823i   &  6.62\\
21 (25)  & 0.15   &  1.01   &  0.520 -- 0.870i     &  6.09\\
23 (31)  & 0.02  & 1.03    &  -0.169 -- 1.012i     & 3.62 \\[0.5ex]
\hline
\normalsize 
\end{tabular} 
\caption{The first $12$ leading Koopman modes estimated with the SPDMD method for the data on velocity magnitude. 
The modes are ordered by their amplitudes. ``Mode's label" $i$ refers to the captured mode with the SPDMD method (with the original DMD index), and ``Period"  $T_{p}={2\pi}/{\mathrm{Im}(\log\lambda_{i})}$ defines the oscillation period in SCALE simulation with $h=0.5$ minutes each (e.g., see Table~\ref{tab:simulation_summary}), calculated from the imaginary component of the eigenvalue in terms of angular frequency.
Besides, the modes with no oscillations are assigned an infinite period ``Inf".}
\label{tabel:scalar-mode-norm-periodic-84} 
\end{table}

\begin{figure}[t] 
\centering
\includegraphics[width=\linewidth]{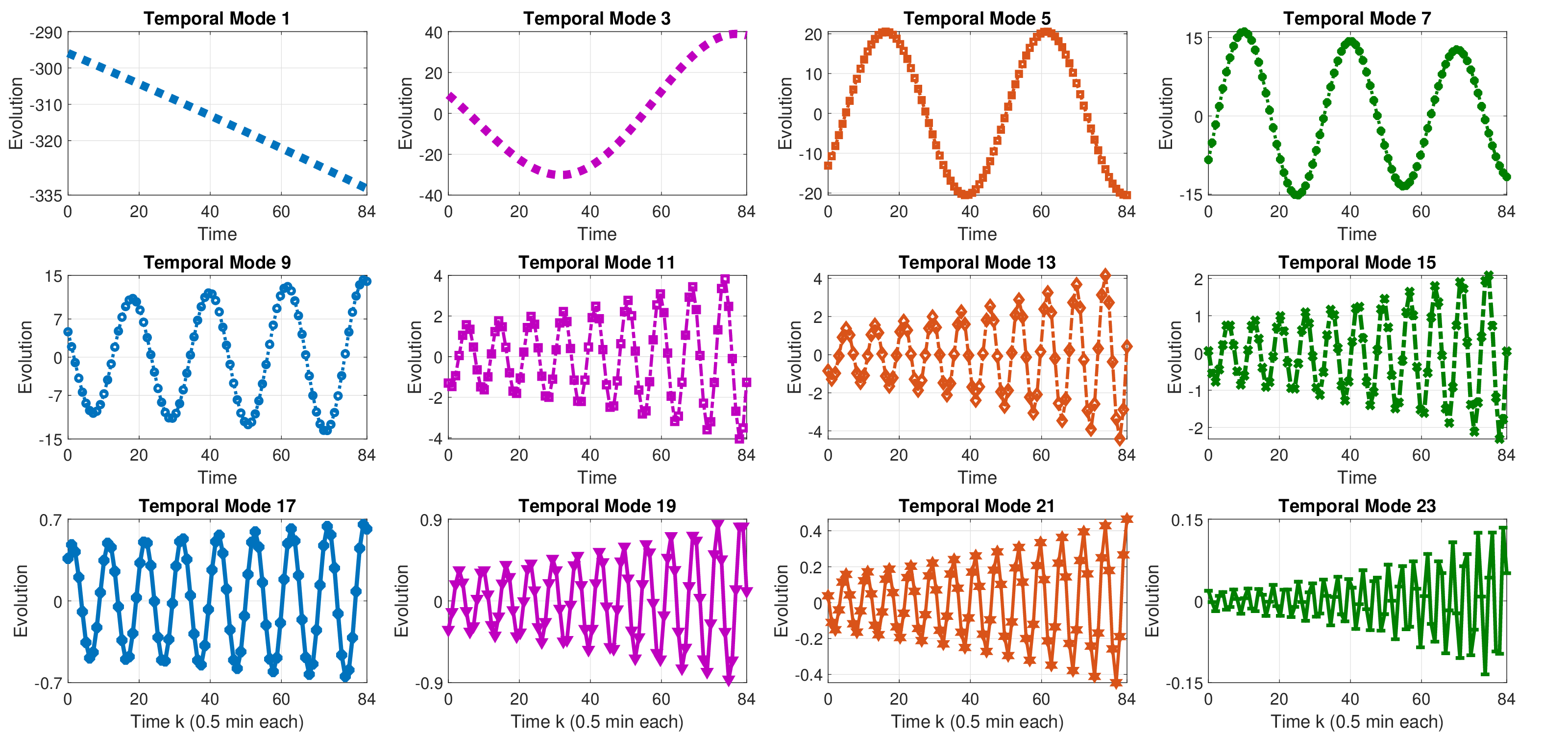}
\caption{Normal evolution~\eqref{eq:normal-evolution} for the SPDMD-based selection in Table~\ref{tabel:scalar-mode-norm-periodic-84}.}
\label{fig:Ex-0-scalar-temporal-dynamics}
\end{figure}

\begin{figure*}[h!] 
\centering
 \includegraphics[width=\linewidth]{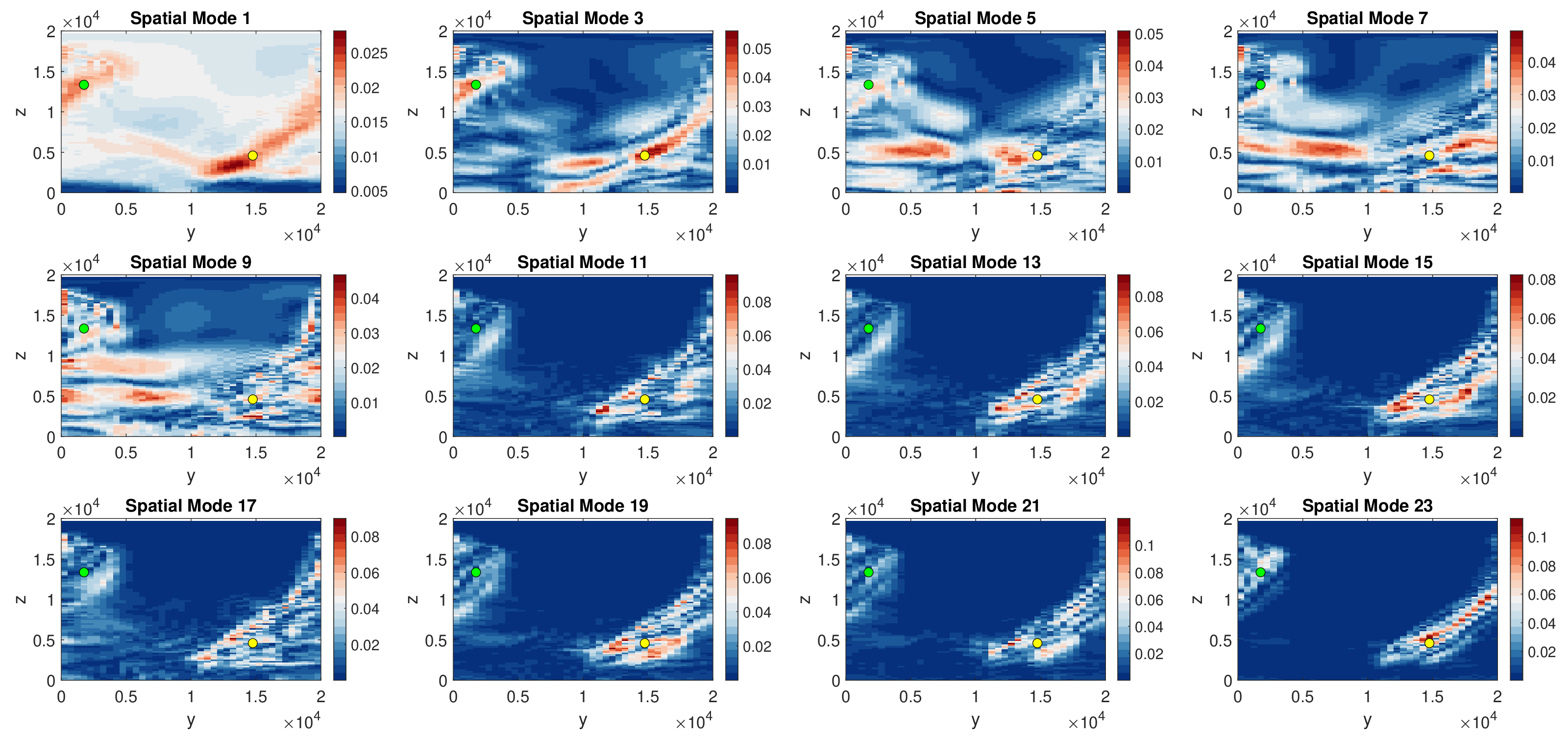}
\caption{Spatial patterns for the first $12$ leading Koopman modes selected using SPDMD. The element-wise absolute values of the Koopman modes ${\bm \phi}_j$ (i.e., the magnitudes $|{\bm \phi}_j|$) for $j=1,3,\cdots,23$ are plotted. In particular, the specific locations marked as the green point are at nearly $
(2~{\rm km},18.1~{\rm km})$ and the yellow point at nearly $(15~{\rm km}, 6.8~{\rm km})$ are highlighted in the $y$–$z$ plane.}
\label{fig:Ex-0-scalar-spatial-modes}
\end{figure*} 

\begin{figure*}[h!]
    \centering
    \includegraphics[width=0.8\linewidth]{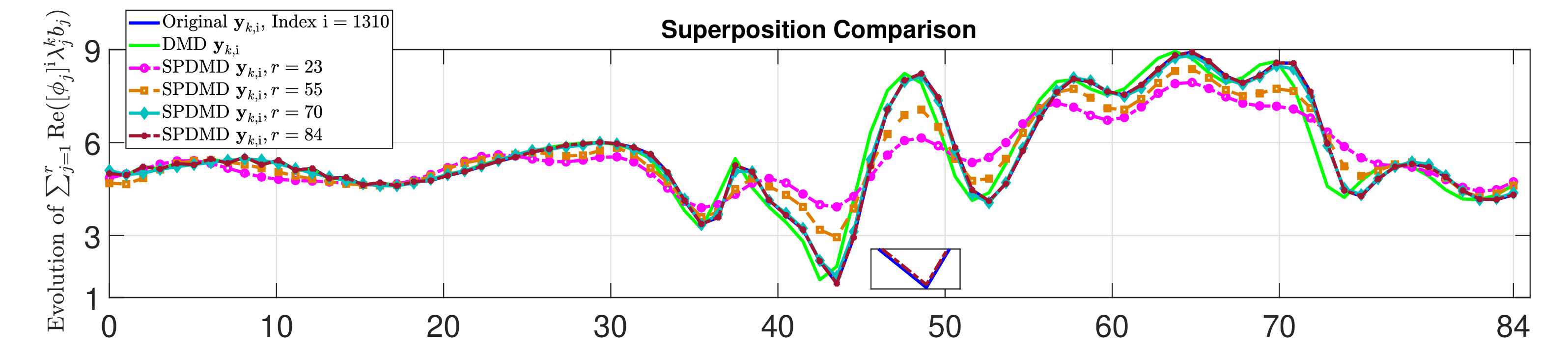}
      \includegraphics[width=0.8\linewidth]{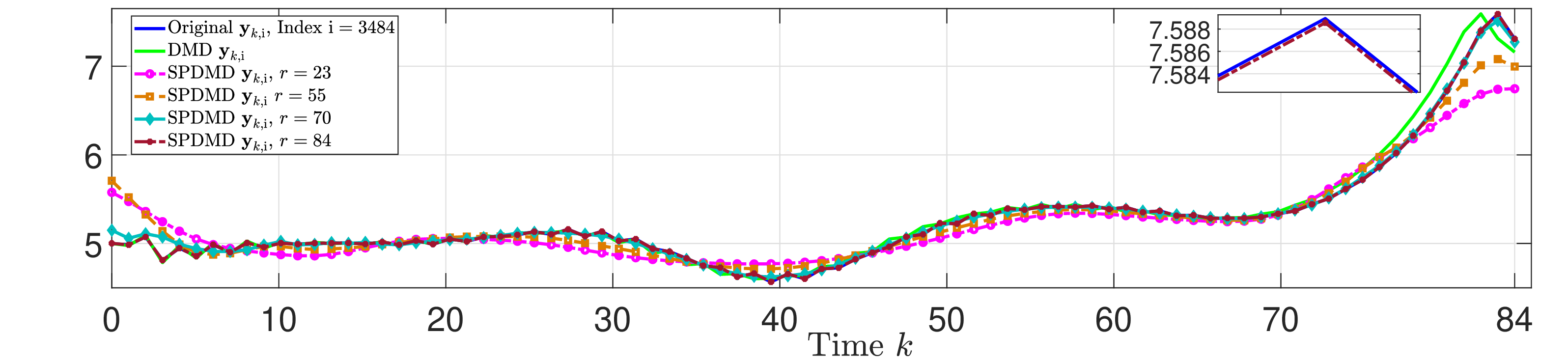}
    \caption{Superposition comparisons \eqref{eq:specific-loaction-superposition} at specific spatial locations (marked by the green and yellow points in Fig.~\ref{fig:Ex-0-scalar-spatial-modes}, respectively) between the original data, full-mode DMD, and SPDMD with varying numbers of selected modes (i.e., $r = 23, 55, 70, 84$).}
    \label{fig:ex-velocity-superposition}
\end{figure*}

To see in detail the selected eigenvalues in Table~\ref{tabel:scalar-mode-norm-periodic-84}, 
we present the normal evolution \eqref{eq:normal-evolution} for the SPDMD-based selection in Fig.~\ref{fig:Ex-0-scalar-temporal-dynamics}.
The first temporal mode shows the evolution of Mode's label~1, which is almost constant in the time duration 
and thus regarded as a \emph{time-average} mode according to $\lambda_{1}\approx1.000+0.000\rm{i}$. 
The remaining extracted temporal modes illustrate time evolutions, all of which are \emph{transient} modes exhibiting either \emph{growth or decay} behavior, 
as listed in Table~\ref{tabel:scalar-mode-norm-periodic-84}. In particular, the seventh mode, with $|\lambda_{7}| \approx 0.99 < 1$, 
represents a \emph{decaying} mode characterized by a mild oscillatory period, where the evolution exhibits gentle oscillations and gradually decays over time. Meanwhile, the other modes with modulus $|\lambda_{i}| > 1$ as shown in Table~\ref{tabel:scalar-mode-norm-periodic-84}, correspond to \emph{growing} modes, whose evolutions oscillate at different slow and fast rates.

Associated with the normal evolutions quantified by the eigenvalues, the Koopman modes ${\bm \phi}_j$ present their spatial patterns as shown in Fig.~\ref{fig:Ex-0-scalar-spatial-modes}. 
Here, the element-wise absolute values of ${\bm \phi}_j$ are plotted. 
By comparison with the raw snapshots of original dynamics in Fig.~\ref{fig:Ex-0-velocity-case-original-data} (or refer to the accessible movie in supplementary material),
the first and third modes directly capture the warm bubble-like patterns with transiently coherent dynamics. However, the first five extracted modes exhibit varying degrees of incoherence, as they have large amplitudes and dominate the transient weather dynamics during the time interval $(30~\mathrm{min}, 42~\mathrm{min})$ in the SCALE simulations.
In addition, the incoherent structures might be related to the turbulent components in the original SCALE simulation, which analysis is part of our future work, connected to the continuous spectra of the Koopman operator \cite{navarra2024variability,colbrook2024rigorous}. 
Interestingly, the remaining spatial modes $|{\bm\phi}_j|$ for $j=11,13,\ldots,23$ still exhibit clear bubble-like structures, despite their relatively small amplitude contributions. These modes are so-called transient \emph{growth} modes, which may be more closely associated with the onset or potential development of heavy rainfall, as shown in Fig.~\ref{fig:Ex-0-scalar-spatial-modes}. In particular, the bubble-like structures emerge at specific spatial locations in the $y$–$z$ plane: $(y_4,z_{88})$, roughly positioned at $(2~\mathrm{km}, 18.1~\mathrm{km})$, is marked as a \emph{green} point; and $(y_{30},z_{33})$, located nearly $(15~\mathrm{km}, 6.8~\mathrm{km})$, is marked as a \emph{yellow} point.
Fig.~\ref{fig:ex-velocity-superposition} compares the original data alongside its reconstruction using DMD and SPDMD methods, based on the superposition evolution \eqref{eq:specific-loaction-superposition}, computed using SPDMD with different numbers of selected modes. For illustration, we evaluate the construction at the green point (indexed as ${\rm i} = 3484$) and the yellow point (indexed as ${\rm i} = 1310$), corresponding to the $\mathrm{i}$-th element of the original data $\mathbf{y}_{k,{\rm i}}$. The superposition results reveal that increasing the number of selected modes$-$specifically $r = 23, 55, 70, 84$$-$progressively improves the reconstruction accuracy. 
It is evident that when all modes ($r = 84$) are selected, the superposition closely reconstructs the original data. In contrast, the superposition retrieved from the less eigenvalues (e.g., $r = 23$) or even fewer modes in Fig.~\ref{fig:ex-velocity-superposition} may fail to track sharp transitions or peaks in the raw data, possibly due to the filtering of important components, such as those continuous spectrum features, which are difficult to capture using a sparse subset of eigenmodes.

\begin{figure*}[t!]
    \centering
    \includegraphics[width=0.7\linewidth]{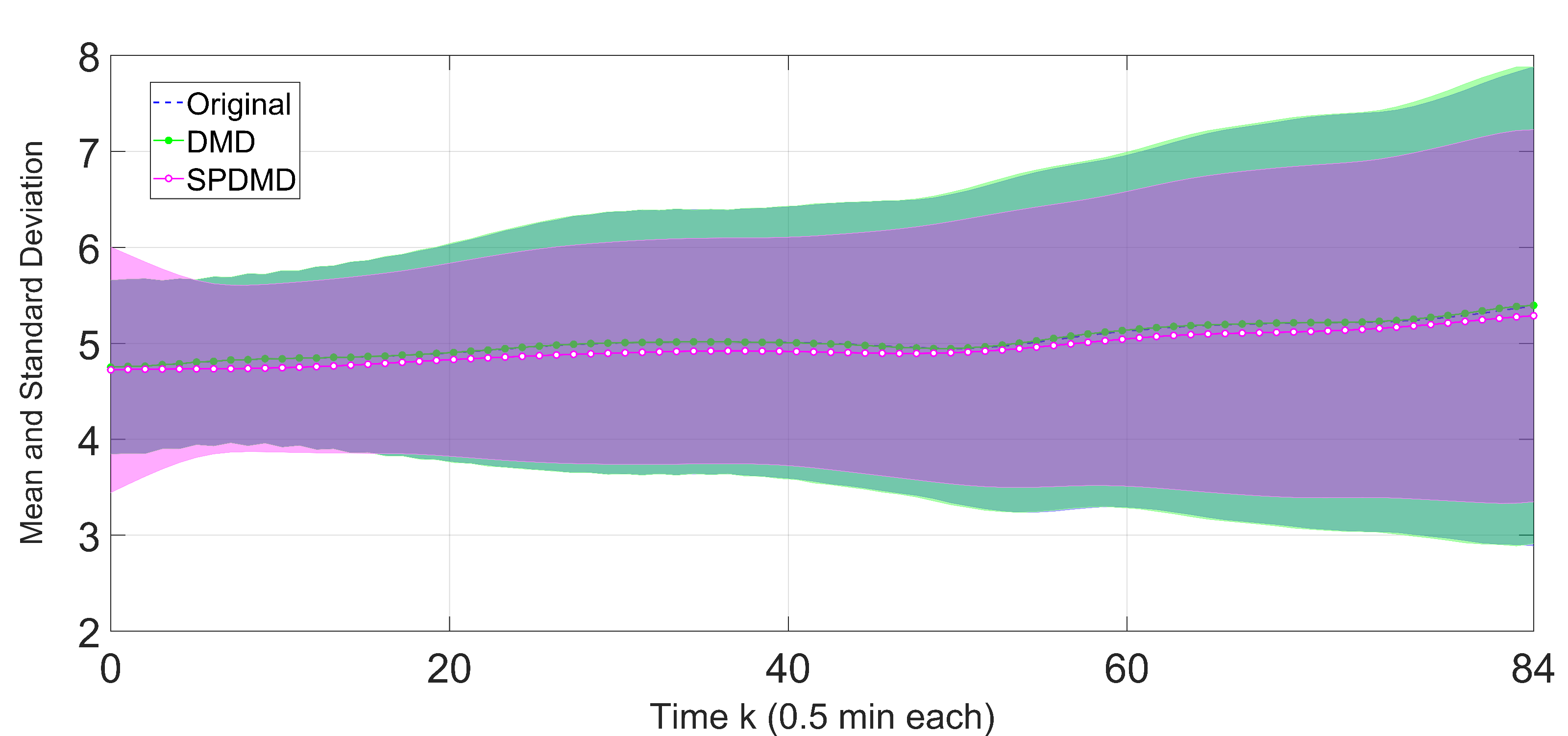}
    \caption{MSD comparison \eqref{eq:mean-standard-der} between the original data (blue), full-mode DMD (green), and SPDMD (magenta) using $r=23$ selected modes. The $x$-axis indicates the run time, and the $y$-axis shows the mean and standard deviation, with colored lines representing the average and shaded areas denoting the full range of MSD across runs for each method applied to the velocity data.}
    \label{fig:msd-velocity-magnitude-84}
\end{figure*}

\begin{figure}[h!] 
\centering
\includegraphics[width=0.7\linewidth]{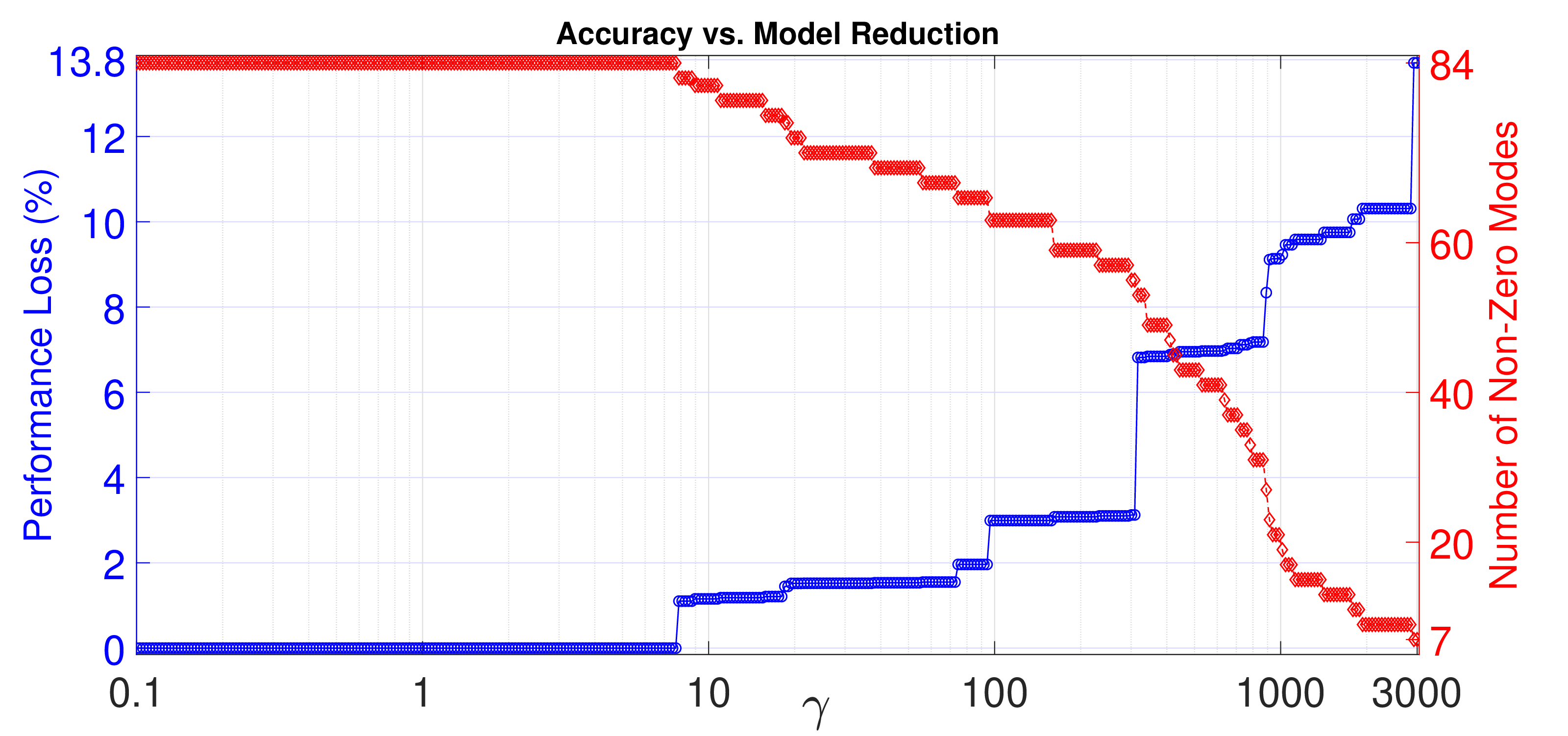}
\caption{A batch of sparsity weights $\gamma\in[0.1,3000]$  with $400$ grid resolutions related to the path of $\bfb_r^{\star}({\gamma})$ the performance loss $\Pi\%$ of the vector of
amplitudes $\bfb_r$ resulting from the SPDMD.}
\label{fig:Ex-0-scalar-amps-weight-loss}
\end{figure} 

An alternative way to quantify differences among the original data, DMD, and SPDMD reconstructions, without focusing on a single spatial location, is to compute the spatial mean and standard deviation (MSD) over the entire $y$–$z$ plane at each time step $k$, for all ${\rm i}\in\{1,\ldots,p\}$:
\begin{align}
    \bar{\bfy}_{k} = \frac{1}{p}\sum_{\rm i=1}^{p}\bfy_{k,\rm i}, \quad 
    {\bm\sigma}_{k} = \sqrt{\frac{1}{p-1}\sum_{\rm i=1}^{p}\left(\bfy_{k,\rm i}-\bar{\bfy}_{k}\right)^{2}},
    \label{eq:mean-standard-der}
\end{align}
where $\bfy_{k,\rm i}$ denotes the $\rm i$-th spatial component of the flattened state vector $\bfy_k\in\mathbb{R}^{p}$ at time snapshot $k=0,1,\ldots,N-1$.
For the reconstructed data (which may contain complex values), only the real part is considered in the computation, e.g., see~\eqref{eq:eigen-decomposition-DMD}.
From Fig.~\ref{fig:msd-velocity-magnitude-84}, an expected full-mode DMD reconstruction (mean: green dotted line; MSD: shaded green area) accurately recovers the statistical evolution of the original data (mean: blue dashed line; MSD: shaded blue area). When SPDMD adopts $r = 23$ modes (mean: magenta line with circle markers; MSD: shaded magenta area), the MSD of SPDMD reconstruction still reflects similar trends of DMD and raw data, albeit with reduced accuracy compared to using a larger number of modes, as also illustrated in Fig.~\ref{fig:msd-velocity-magnitude-84}. Although some residual error remains, the results are of remarkably good recovery capability, considering the selected few modes in the latent space determined by the large sparsity penalty, and the choice of observables governs the raw data. We observed that the sparsity weight $\gamma$ and the quantity of measured data play a critical role in determining the density of the MSD, which is essential for effective model reduction. A comprehensive discussion of the accuracy and model-complexity plots across a range of sparsity-weighting variables is provided in the following context. In a nutshell, SPDMD has been shown to be effective at extracting dominant transient modes from the velocity magnitude data field. Moreover, this experiment suggests that capturing such bubble-like structures more appropriately may require further investigation using relatively shorter-term SCALE simulations, which will be explored in the next Subsection~\ref{subsec:vorticity-magnitude}.

In what follows, we quantify the selection of the estimated eigenvalues with SPDMD. 
For this, let us define the so-called performance loss, that is, the error $J_0(\bfb_r^{\star})=\|\bfY-{\bf\Phi}_r\mathrm{diag}(\bfb_r^{\star})\bfT_r\|_{\text{F}}^{2}$ evaluated at the sparse optimal amplitudes $\bfb_r^{\star}$, which can be obtained via the closed-form solution \eqref{eq:QP-closed-form-solution}, and normalized with the baseline $J_0(\mathbf{0}) = \|\bfY\|_{\mathrm{F}}^2$, giving rise to the relative loss metric:
\begin{align*}
  \Pi\% =\sqrt{\frac{J_0(\bfb_r^{\star})}{J_0({\mathbf{0})}}}\times{100}=
  \frac{\|\bfY-\mathbf{\Phi}_r\mathrm{diag}(\bfb_r^{\star})\bfT_r\|_{\mathrm{F}}^{2}}{\|\bfY\|_{\rm{F}}^{2}}\times{100}.
\end{align*}
The case $\Pi\%=0$ implies no loss of the modeling by the DMD method, while the case $\Pi\%=100$ implies another extreme case where no Koopman mode is selected. 
As we impose a greater emphasis on the sparsity weight $\gamma$ (resp., the sparsity level of amplitudes), the number of non-zero elements in the vector $\bfb_r$, called the number of amplitude of $\bfb_r^{\star}$, decreases, leading to a reduction in the quality of the least-squares approximation, we refer to see Remark~\ref{remark:role-of-sparsity} for the similar statement. 
This is indeed shown in Fig.~\ref{fig:Ex-0-scalar-amps-weight-loss}, where we take the range of sparsity weight as $\gamma\in[0.1,3000]$. 
The condition $\gamma_{\min}=0.1$ produces a dense amplitudes $\bfb_r^{\star}$ with $84$ non-zero elements, while $\gamma_{\max}=3000$ yields a highly sparse solution $\bfb_r^{\star}$ containing only $7$ non-zero elements. 
The sparsity level, measured by the number of non-zero amplitudes (i.e., amplitude cardinality), and the performance loss $\Pi\%$ for the optimal amplitude vector $\bfb_r^{\star}$ derived from SPDMD are illustrated in the right figure of Fig.~\ref{fig:Ex-0-scalar-amps-weight-loss}. 
This relationship is shown as a function $\bfb_{r}(\gamma)$ of the user-specified parameter $\gamma\geq0$, which balances the trade-off between approximation quality and solution sparsity. 
As expected, increasing $\gamma$ results in sparser solutions of amplitudes but compromises the quality of the least-squares approximation.

\subsection{Vorticity Magnitude}\label{subsec:vorticity-magnitude}
Without loss of generality, we consider an alternative choice of observables, namely data on vorticity magnitude, to quantify the time-dependent warm, bubble-like patterns, as suggested by the data source described in Subsection~\ref{subsec:data-source}. Partially selected snapshots of the original data are shown in Fig.~\ref{fig:Ex-1-Vorticity-case-original-data}.
In the following, we address the two cases as our analyses: Case~1 for the initial stage where the bubble-like pattern emerges from the earth's surface; Case~2 for the mid-term stage where the bubble-like pattern tends to be resolved.

\begin{figure*}[!t] 
\centering
\includegraphics[width=0.4\linewidth]{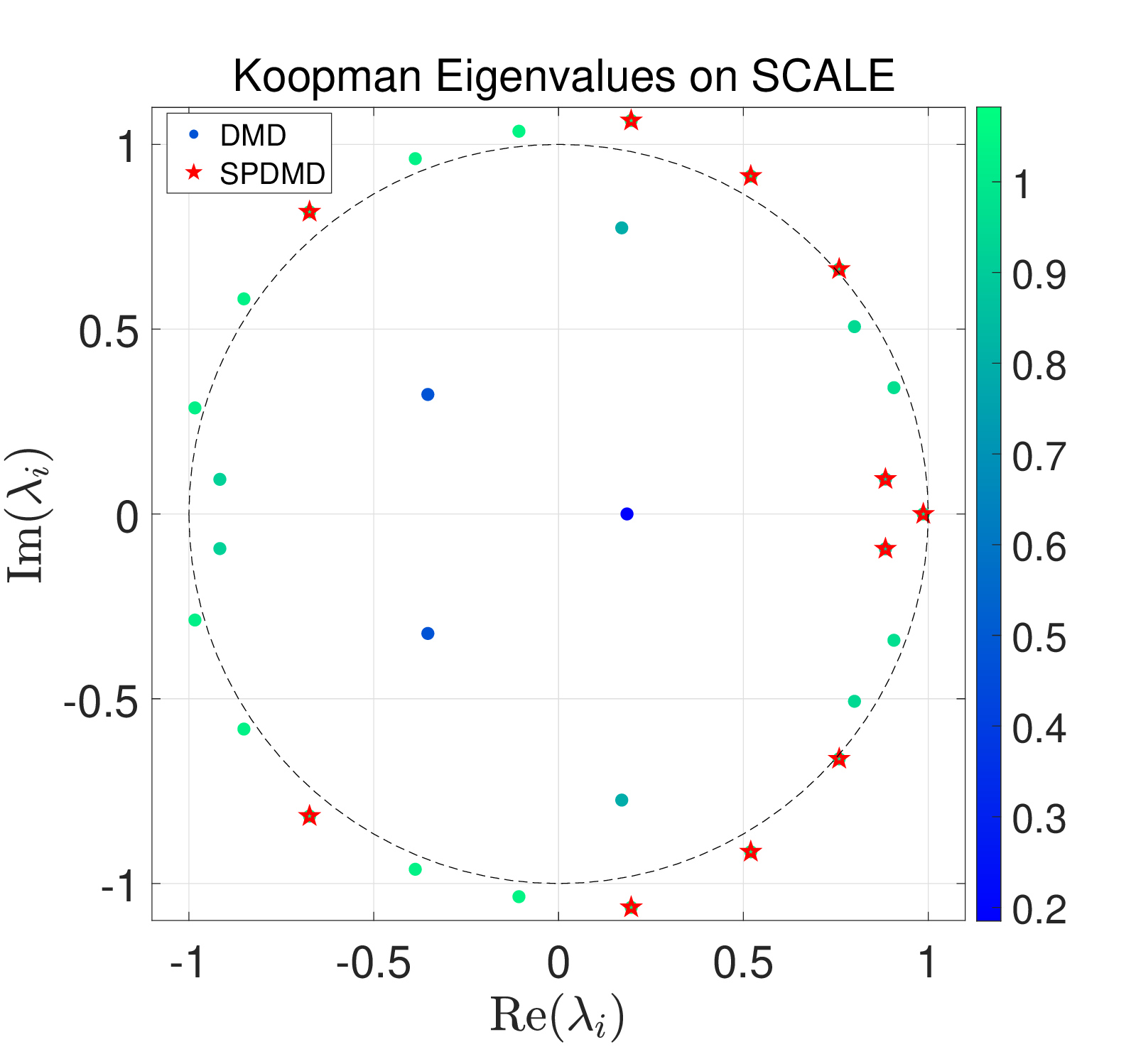}\quad
\includegraphics[width=0.4\linewidth]{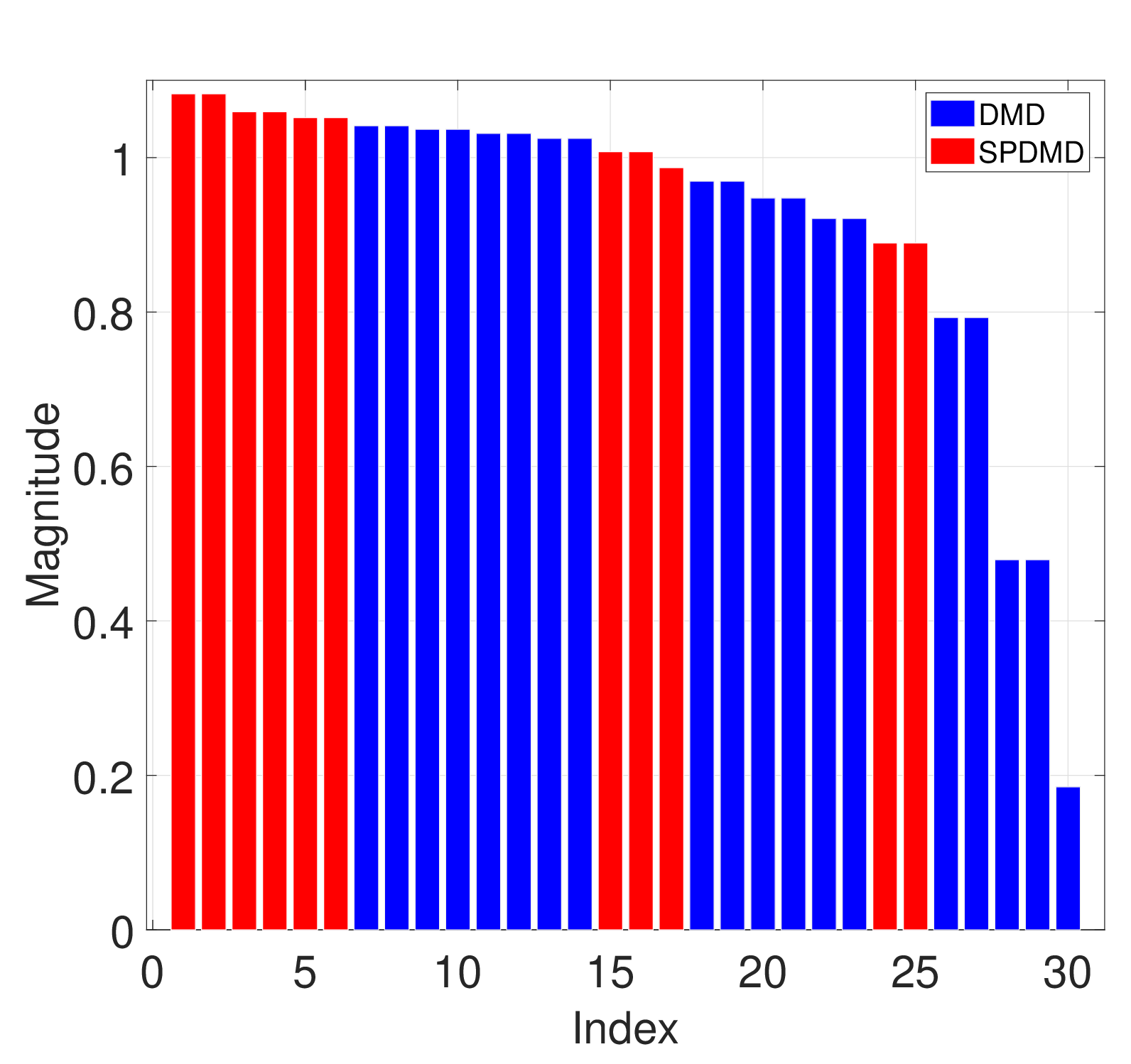}
\caption{Distribution and absolute values of Koopman eigenvalues estimated with the DMD and SPDMD methods for the data on vorticity magnitude: Case~1.}
\label{fig:vorticity-eigenvalues}
\end{figure*} 

\begin{table}[t] 
\centering 
\caption{The first $6$ dominant transient Koopman modes selected by the SPDMD method for the data on vorticity amplitude: Case~1.}
\label{tabel:vortex-mode-norm-periodic-v30} 
\begin{tabular}{@{\extracolsep{5pt}} c c c c c} 
\\[-1.8ex]\hline 
\hline \\[-1.8ex]
 Mode's label & DMD Amps. & Absolute value & Eigenvalue & Period ($0.5$ min)\\ 
  $i$  & $|b_{i}|$ & $|\lambda_{i}|$ & $\lambda_{i}$ & ${2\pi}/{\mathrm{Im}(\mathrm{In}\lambda_{i})}$\\
\hline \\[-1.8ex] 
 \rowcolor{lightgray}
1 (1)  & 7.08   & 0.99   & 0.987 + 0.000i  &     Inf \\
 \rowcolor{lightgray}
3 (3)  &  1.31  & 0.89  &  0.884 -- 0.094i  &  59.16 \\

5 (11)  & 0.14  & 1.05 &  0.520 -- 0.914i   &   5.96 \\

7 (13)   & 0.11 & 1.08 &  0.196 -- 1.065i    & 4.53 \\

9 (7)   & 0.10   &  1.01  &  0.759 -- 0.662i  &  8.76 \\
11 (21)  & 0.01 & 1.06 & -0.674 -- 0.818i &  2.78 \\ [0.5ex]
\hline
\normalsize 
\end{tabular} 
\end{table}

\subsubsection{Case 1}\label{experiment-vorticity-case-1}
First, we consider a relatively short-term SCALE weather experiment with a duration of $T_s=15.5$ minutes and a time resolution of $h=0.5$ minutes, yielding $ N=T_s/h=31$ snapshots of the vorticity magnitude (i.e., $k=0,1,\ldots,30$). We then present the spectral analysis for the corresponding data matrix $\bfY\in\mathbb{R}^{3880\times{31}}$.
The DMD and SPDMD methods are applied to the data matrix, and the estimated eigenvalues are presented in Fig.~\ref{fig:vorticity-eigenvalues} and Table~\ref{tabel:vortex-mode-norm-periodic-v30}.  
The left side of Fig.~\ref{fig:vorticity-eigenvalues} shows the distribution of the estimated eigenvalues in the complex plane, while the right figure displays their magnitudes (i.e., absolute values) sorted in descending order. The eigenvalues determined by the SPDMD method ($11$ modes, roughly forming $6$ conjugate pairs) are marked with \emph{red stars}, in contrast to the \emph{blue points} corresponding to those obtained from standard DMD.

\begin{figure}[t!]
\centering
\includegraphics[width=\linewidth]{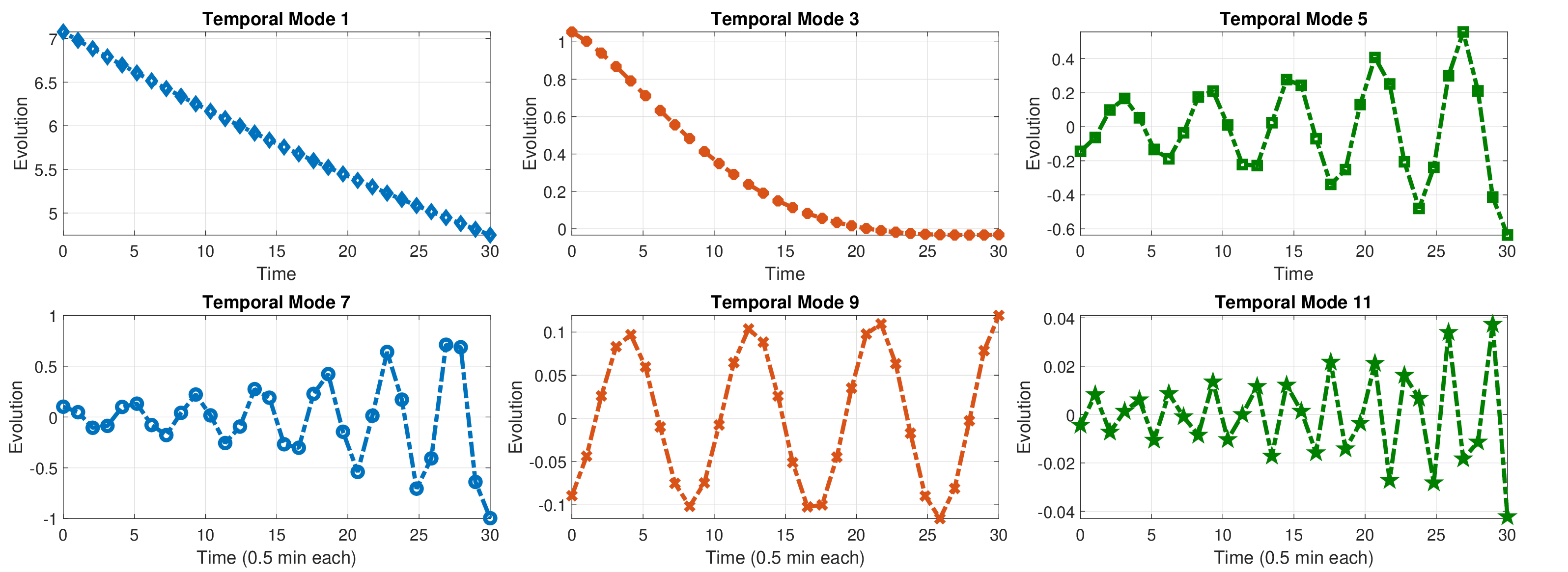}
\caption{Normal evolution~\eqref{eq:normal-evolution} for the SPDMD-based selection in Table~\ref{tabel:vortex-mode-norm-periodic-v30}.}
\label{fig:temporal-vorticity-v30}
\end{figure}

\begin{figure}[t!]
\centering
\includegraphics[width=\linewidth]{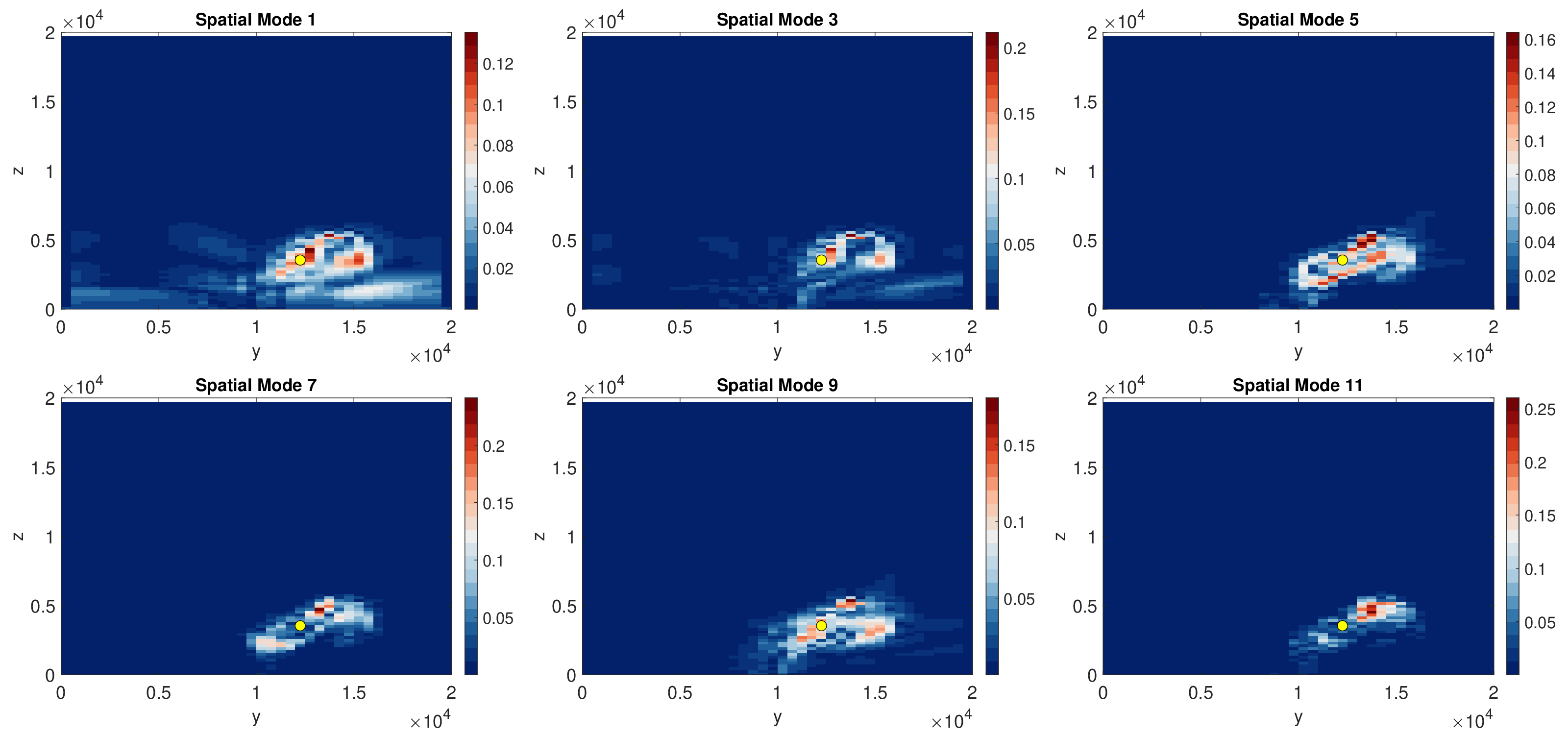}
\caption{Spatial patterns of the Koopman modes selected via SPDMD: Case~1, illustrating the emergence of a warm-bubble-like pattern. The element-wise absolute values of the Koopman modes $\bm{\phi}_j$ are visualized, with a specific spatial location near $(12.5{\rm km}, 5.4{\rm km})$ marked as a yellow point.}
\label{fig:vorticity-spatial-modes-V30}
\end{figure}

Associated with the eigenvalues, Fig.~\ref{fig:temporal-vorticity-v30} shows the normal evolution \eqref{eq:normal-evolution} by SPDMD-selected method with $r=11$. 
Meanwhile, the estimated Koopman modes $\bm{\phi}_j$ are presented in Fig.~\ref{fig:vorticity-spatial-modes-V30}, where the element-wise absolute values (that is, $|\bm\phi_{j}|$ for $j=1,3,5,7,9,11$) in Fig.~\ref{fig:vorticity-eigenvalues} are visualized. Obviously, the first two modes selected by SPDMD have magnitudes less than $1$, indicating decaying behavior.
Physically, these modes reflect the evolving dynamics of the vorticity field following a rainfall event. More precisely, the first mode, $\lambda_1 \approx 0.987 + 0.000{\rm{i}}$ with $|\lambda_1| \approx 0.99$, represents persistent, slowly decaying background moisture or light rainfall. The second mode, $\lambda_2 \approx 0.884 - 0.094{\rm{i}}$ with $|\lambda_2| \approx 0.89$, captures faster-decaying, weak, short-lived convective activity. Together, the first two modes characterize the \emph{decay phase} of precipitation-related dynamics.
However, the remaining four modes, with eigenvalues satisfying $|\lambda_i| > 1$ for $i = 5, 7, 9, 11$, correspond to transiently \emph{growing} features. These modes are particularly relevant for capturing the onset of organized structures, such as the emergence of bubble-like vorticity patterns from the surface. It is significant because it indicates the relatively early stages of warm, bubble-like patterns, which are transiently growing modes indicating the initiation or forthcoming development of heavy convective rainfall.
Needless to say, detecting growth modes can provide valuable insights for designing control protocols to stabilize the original weather system by targeting and regulating these few key unstable modes.
By comparison with the original snapshots in Fig.~\ref{fig:Ex-1-Vorticity-case-original-data}, the spatial modes shown in Fig.~\ref{fig:vorticity-spatial-modes-V30} illustrate that the emergence of warm-bubble patterns during the initial stage is successfully captured by the growing and decaying Koopman modes selected via SPDMD.

\begin{figure}[t!]
\centering
\includegraphics[width=0.65\linewidth]{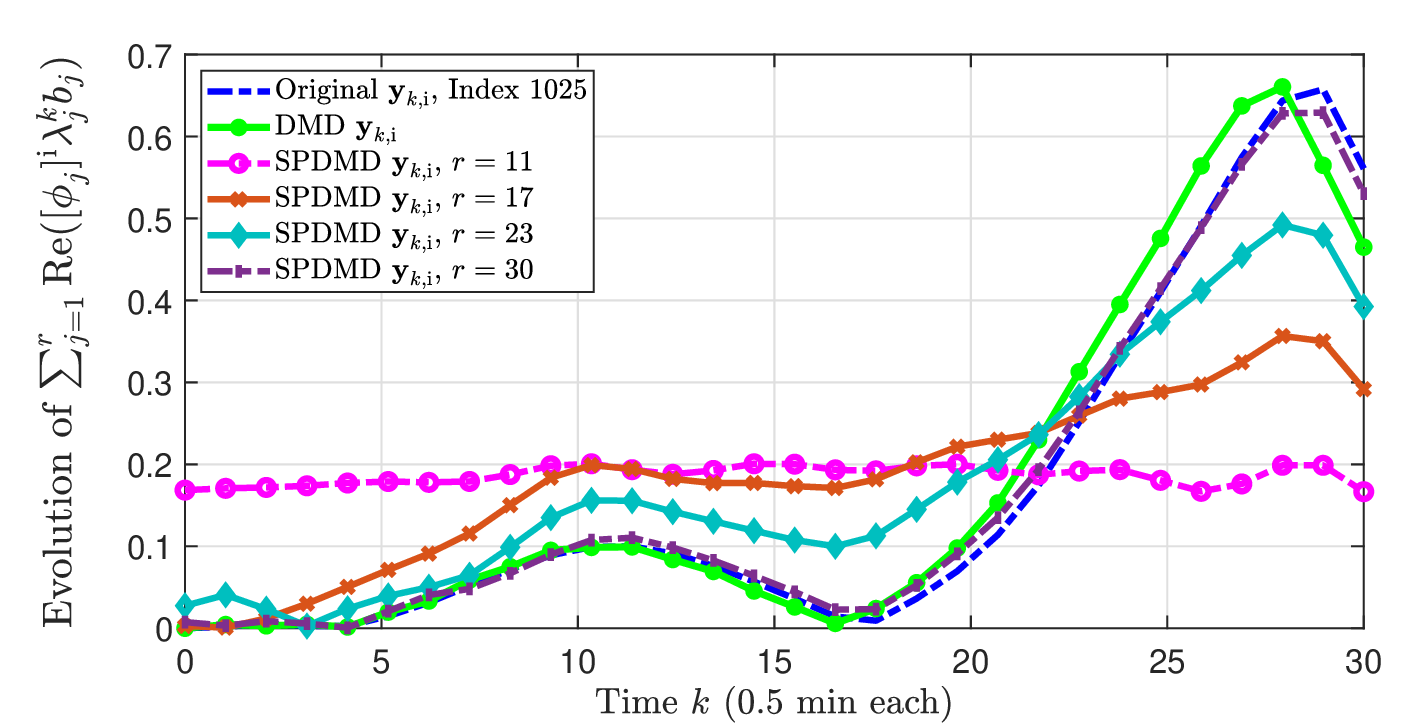}
\includegraphics[width=0.65\linewidth]{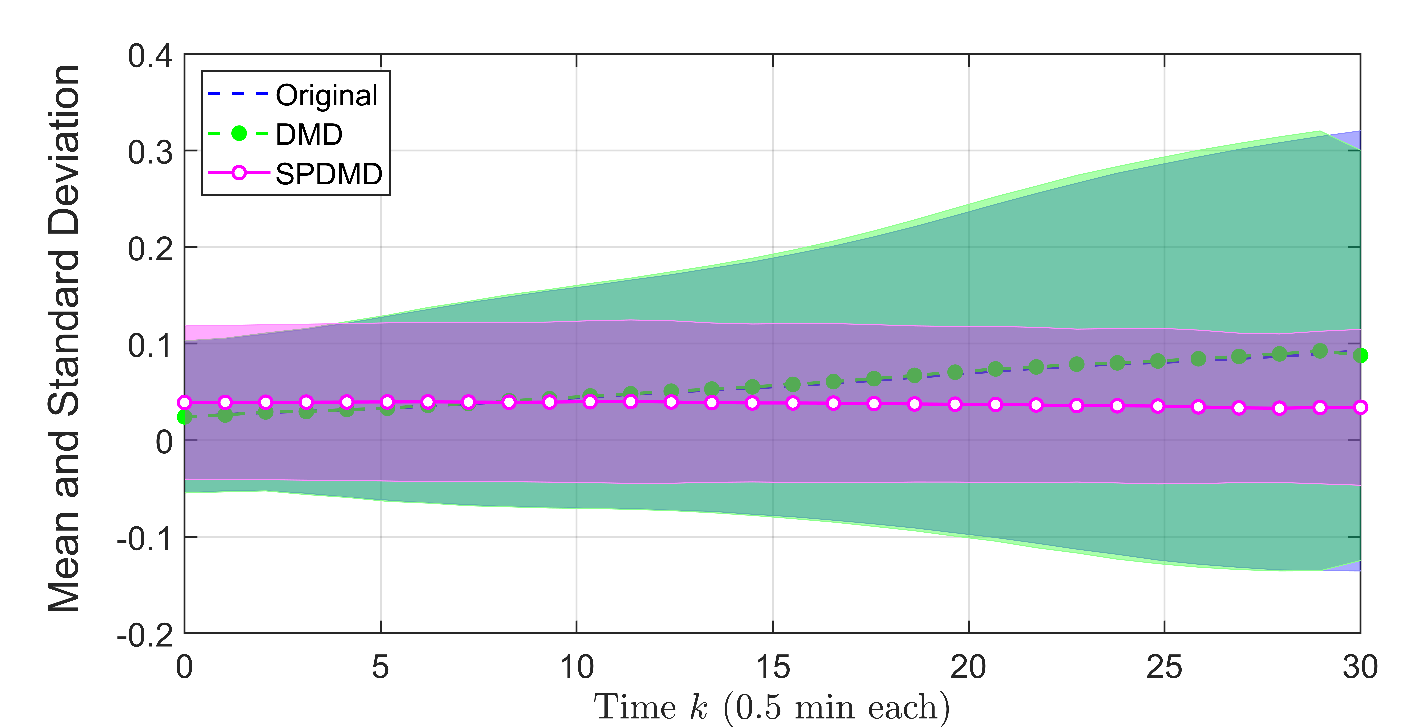}
 \caption{(top): Superposition comparison \eqref{eq:specific-loaction-superposition} at a specific location indicated in Fig.~\ref{fig:vorticity-spatial-modes-V30} between the original data sequence, full-mode DMD, and SPDMD with varying numbers of selected modes ($r = 11, 17, 23, 30$) in the vorticity data fields.
(bottom): The MSD comparison \eqref{eq:mean-standard-der} using SPDMD with $r = 11$ selected modes.}
\label{fig:superposition-MSD-V30}
\end{figure}

\begin{figure}[thb!]
\centering
\includegraphics[width=0.7\linewidth]{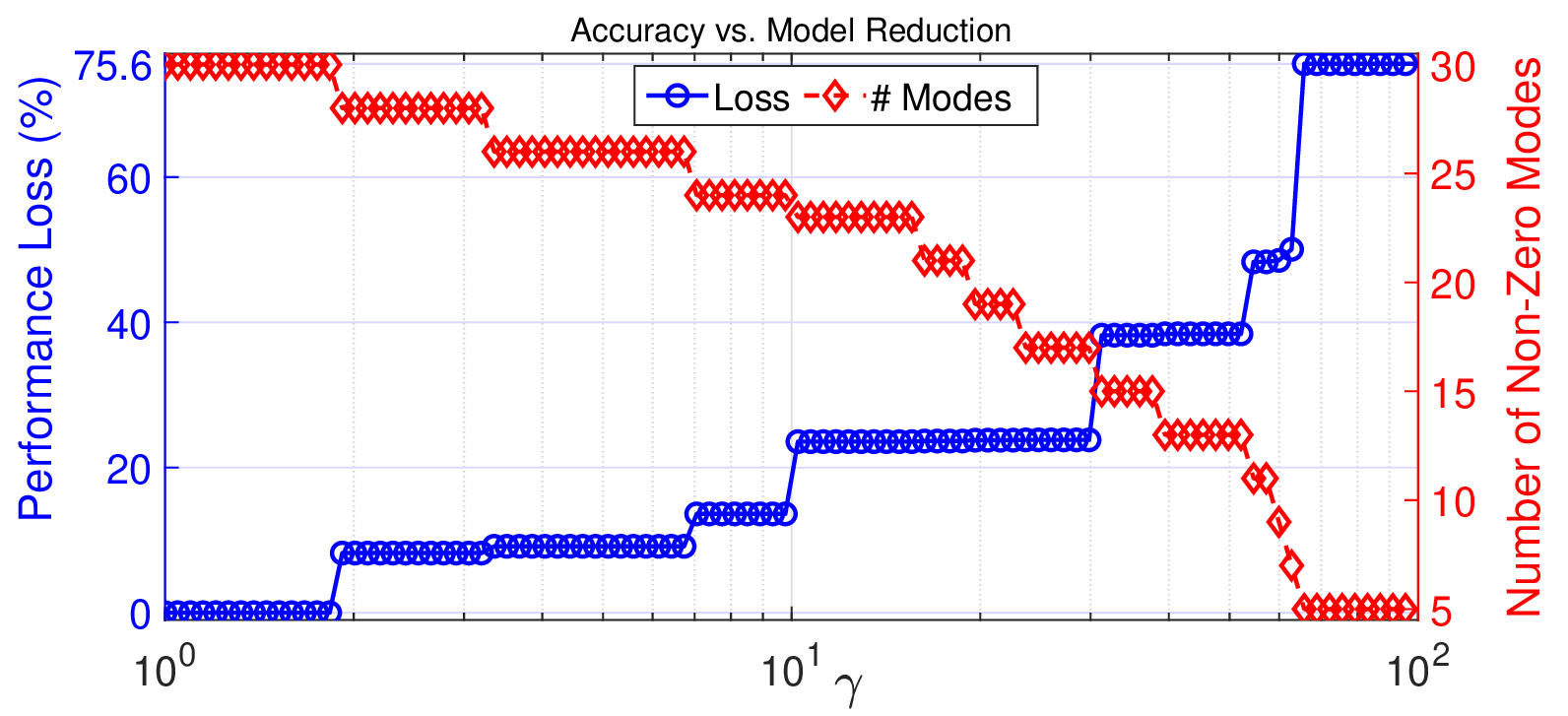}
\caption{The accuracy is estimated by the performance loss relative to the model-reduction, which varies with different weights $\gamma\in[1,100]$ with $350$ grids through SPDMD.}
\label{fig:loss-accuracy-v30}
\end{figure}

To deepen the analysis, we take another look at the comparison between the original data in detail 
and reconstructed data. 
The reconstruction is done by truncating the summation on the right-hand side of \eqref{eq:dynamic-mode-decomposition-1}, giving rise to the real-valued superposition evolution \eqref{eq:specific-loaction-superposition}. 
As shown in the {top portion} Fig.~\ref{fig:superposition-MSD-V30}, the real-valued superposition evolution of $\bfy_{k,\mathrm{i}}$ with the \emph{index} $\mathrm{i} = 1025$ is compared among the original data, full-mode DMD, and SPDMD with varying numbers of selected modes. This $\mathrm{i}$-th component corresponds to the specific spatial location marked by the yellow point at $(y_{25}, z_{26})$ corresponding to approximately $(12.5~\mathrm{km}, 5.4~\mathrm{km})$, as highlighted in Fig.~\ref{fig:vorticity-spatial-modes-V30}. 
While the full-mode DMD method recovers the original data with high accuracy, the SPDMD method requires an increasing number of modes, such as $r = 11, 17, 23, 30$, to progressively improve accuracy and approach the original data results.
Besides, to further quantify the reconstruction effect across the entire $y$–$z$ plane for all data points in the sequence, we compute the MSD \eqref{eq:mean-standard-der}; the corresponding comparisons are shown in the bottom panel of Fig.~\ref{fig:superposition-MSD-V30}.
The legend markers follow the same convention as in previous experiments: the \emph{blue dashed line} represents the mean of the original data, the \emph{green dotted line} corresponds to the DMD-based reconstruction, and the \emph{magenta line with circle markers} denotes the SPDMD-based reconstruction.
From Fig.~\ref{fig:superposition-MSD-V30}, we can see that both the DMD and SPDMD-selected based methods show a relative error or residual compared to the original data. 
This discrepancy arises because we are considering only a short-term SCALE simulation in Case 1 of the vorticity magnitude field, where the original data matrix $\mathbf{Y} \in \mathbb{R}^{3880 \times 31}$ contains rich spatial information across $3880$ grid points but relatively limited temporal information with only $31$ snapshots.

Before concluding {Case~1},
Fig.~\ref{fig:loss-accuracy-v30} presents a cross-validation plot illustrating the trade-off between performance loss and the number of non-zero modes in terms of the sparse amplitude vector $\bfb_r(\gamma)$. The evaluation is conducted over a logarithmically spaced grid of $\gamma$ values, ranging from $\gamma_{\min} = 1$ to $\gamma_{\max} = 100$, with $\gamma_{\text{grid}} = 350$ points.
According to the results in Fig.~\ref{fig:loss-accuracy-v30}, using full modes in the SPDMD method results in nearly zero performance loss, whereas selecting only $5$ modes leads to a significant loss of approximately $\Pi = 75\%$. This demonstrates that the SPDMD method can achieve a favorable balance between accuracy and model-reduction by appropriately tuning the sparsity weight and selecting an adequate number of modes.
Specifically, the SPDMD-selected method effectively captures the warm, bubble-like pattern of the spatial structure, even with a limited number of snapshots and for short-term vorticity data sequences.

\begin{figure*}[t!] 
\centering
\includegraphics[width=0.4\linewidth]{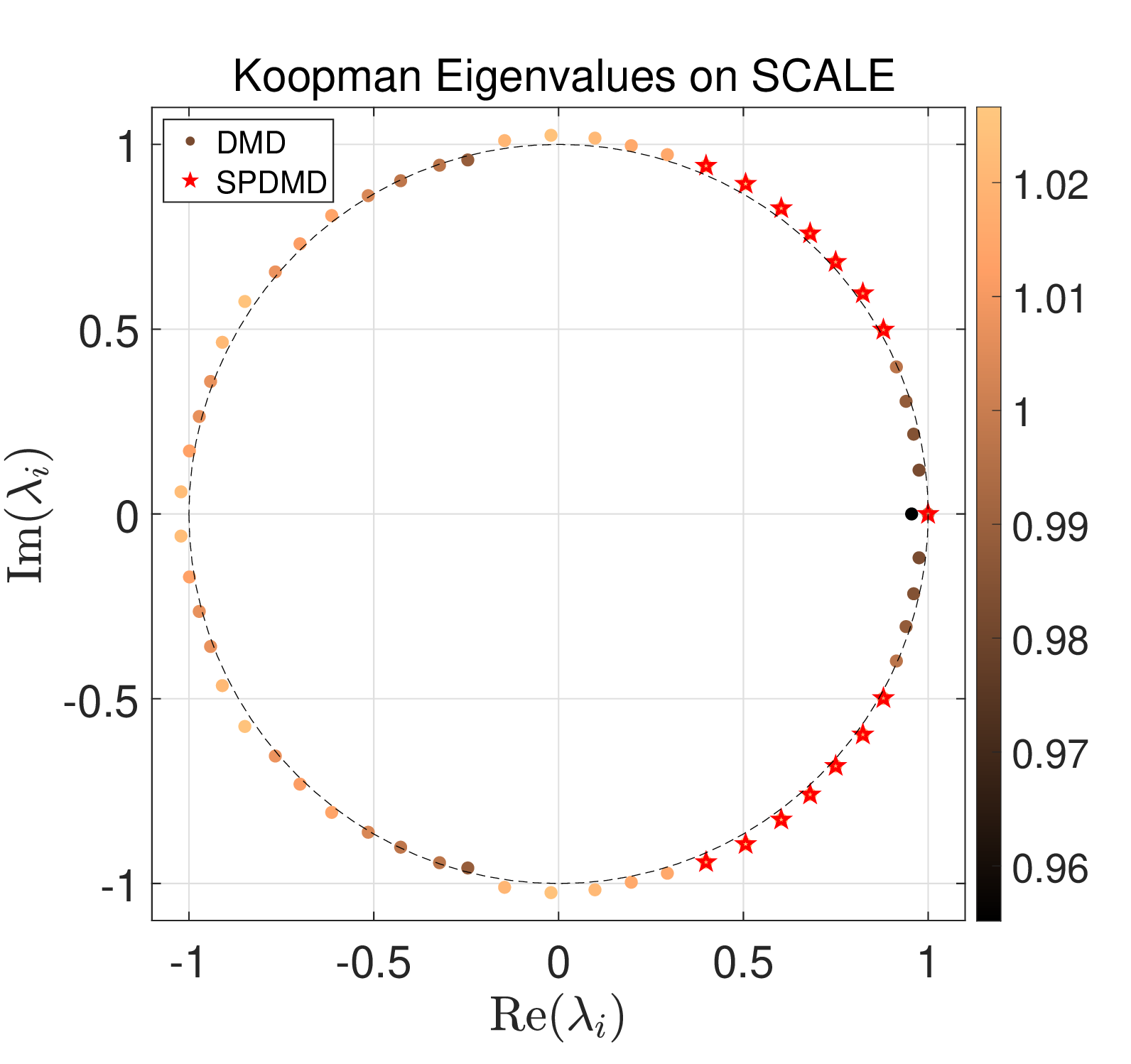}
\includegraphics[width=0.4\linewidth]{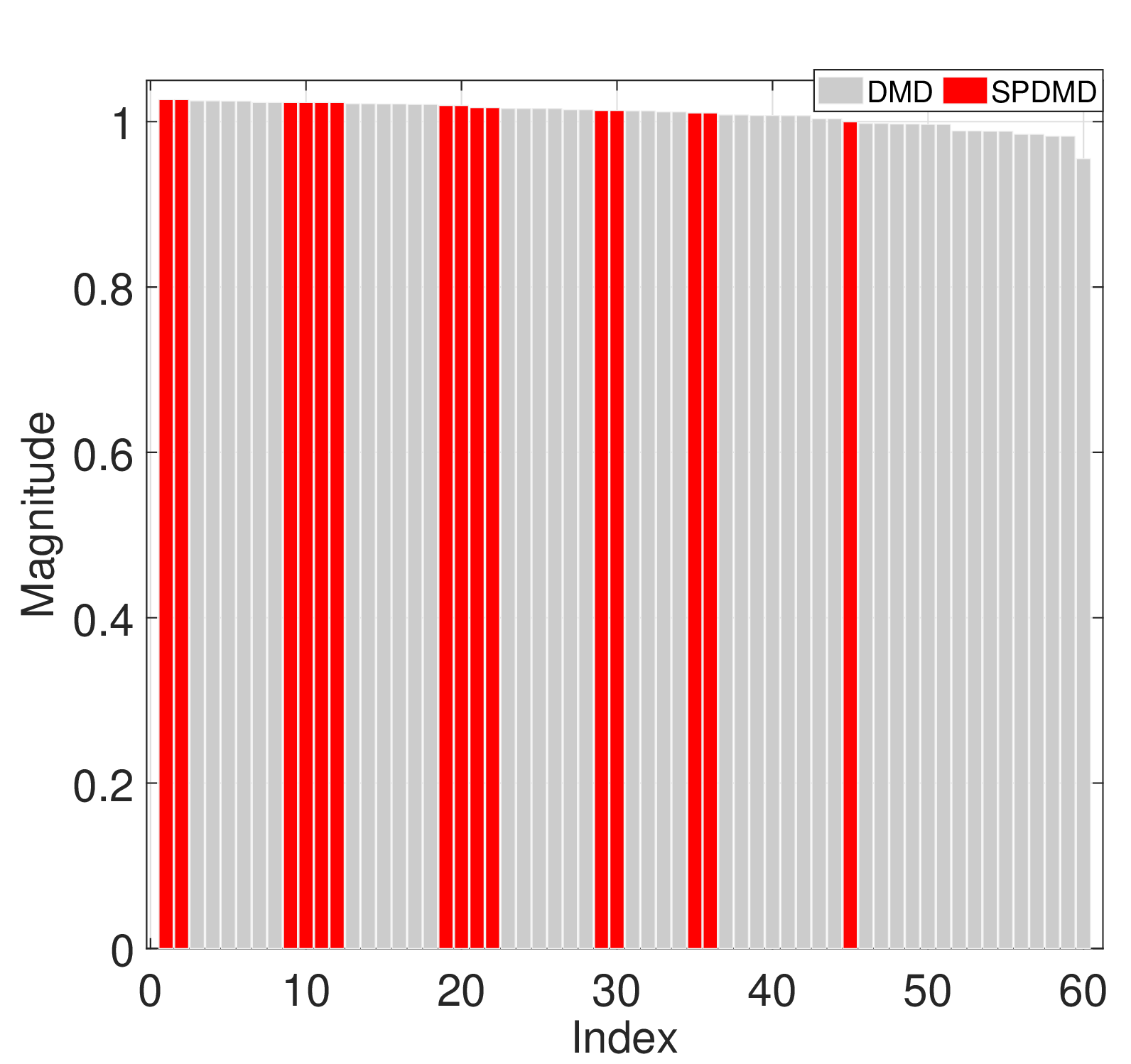}
\caption{Distribution and absolute values of Koopman eigenvalues estimated with the DMD and SPDMD methods for the data on vorticity magnitude: Case 2.}
\label{fig:vorticity-eigenvalues-60}
\end{figure*}

\subsubsection{Case~2}\label{experiment-vorticity-case-2}
Second, we consider a mid-term SCALE weather simulation with a duration of $T_s=30.5$ minutes and a sampling interval $h=0.5$ minutes, which yields $N=T_s/h=61$ snapshots (i.e., $k=0,1,\ldots,61$) of the vorticity magnitude. Following a similar procedure to Case~1, we perform spectral analysis on the resulting data matrix $\mathbf{Y}\in \mathbb{R}^{3880 \times 61}$. However, there are significant differences in the evolution of warm bubble-like patterns across different time stages in weather simulations. As shown in Fig.~\ref{fig:Ex-1-Vorticity-case-original-data}, which displays snapshots generated from raw data at time steps $k = 41, 45, \ldots, 61$ (corresponding to the time duration (20.5 min, 30.5 min) of the weather simulations), the mid-term stage primarily exhibits persistent growth behavior.
The $60$ eigenvalues are estimated with the DMD method, and the principle $15$ eigenvalues are selected with the SPDMD method, as shown in Fig.~\ref{fig:vorticity-eigenvalues-60} and Table~\ref{tabel:vortex-mode-norm-e-folding-periodic-v60}.

\begin{table}[t!] 
\centering 
\caption{
The first $8$ leading Koopman modes selected by the SPDMD method for the data on vorticity
amplitude: Case 2
} 
\label{tabel:vortex-mode-norm-e-folding-periodic-v60} 
\begin{tabular}{@{\extracolsep{5pt}} c c c c c} 
\\[-1.8ex]\hline 
\hline \\[-1.8ex]
 Mode & Amps & Norm & Eigenvalue & Period ($0.5$ min)\\ 
  $i$  & $|b_{i}|$ & $|\lambda_{i}|$ & $\lambda_{i}$ & ${2\pi}/{\mathrm{Im}({\log}\lambda_{i})}$\\
\hline \\[-1.8ex] 
  \rowcolor{lightgray}
1 (1)    &  7.76    &   1.00   &   1.000 + 0.000i    &       Inf \\
3  (12)   &   0.47   &    1.01  &    0.879 -- 0.498i    &      12.18\\
5 (14)    & 0.40     &  1.02    &   0.823 -- 0.597i    &   10.02\\
7 (18)    & 0.21     & 1.02   & 0.681 -- 0.759i   & 7.48 \\
9 (20)    & 0.15   & 1.02  &  0.602 -- 0.827i    & 6.68 \\
11 (16)   &   0.14   &    1.01   &   0.750 -- 0.682i   &       8.51 \\
13 (22)   &  0.12     &  1.03   &   0.506 -- 0.893i     & 5.95\\
15 (24)  &   0.04      & 1.02   &   0.399 -- 0.942i      &      5.37 \\[0.5ex]
\hline
\normalsize 
\end{tabular} 
\end{table}


\begin{figure}[t!]
\centering
\includegraphics[width=\linewidth]{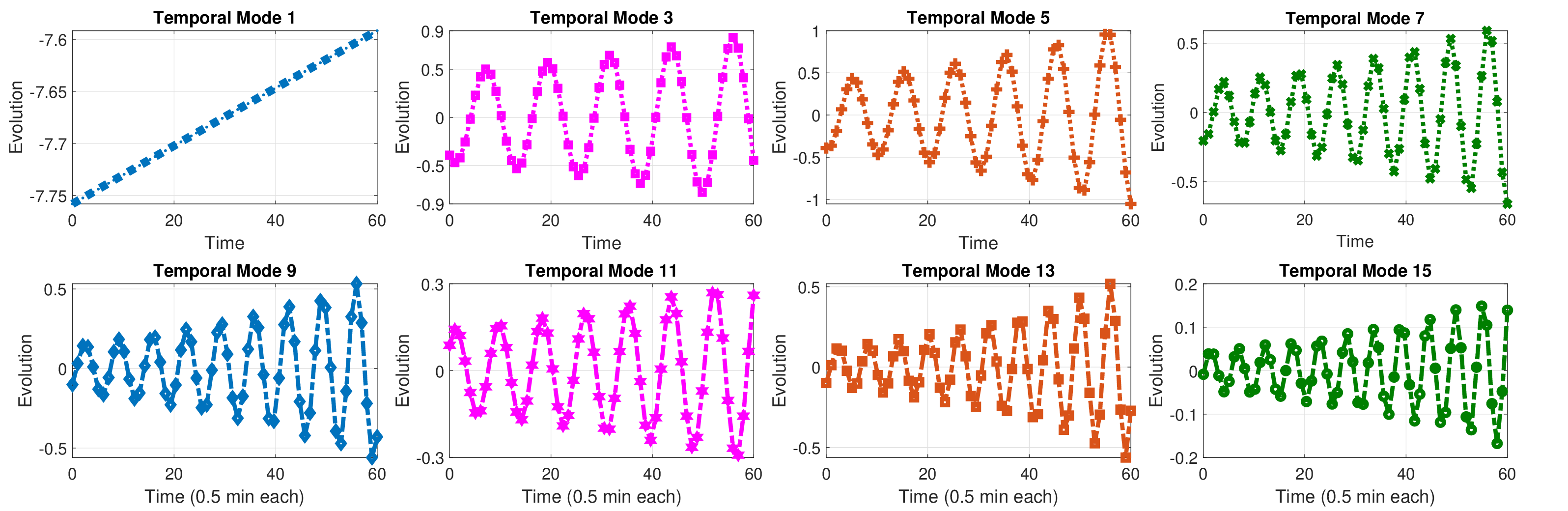}
\caption{
Normal evolution~\eqref{eq:normal-evolution} for the SPDMD-based selection in Table~\ref{tabel:vortex-mode-norm-e-folding-periodic-v60}. 
}
\label{fig:temporal-vorticity-V60}
\end{figure}

\begin{figure}[t!]
\centering
\includegraphics[width=\linewidth]{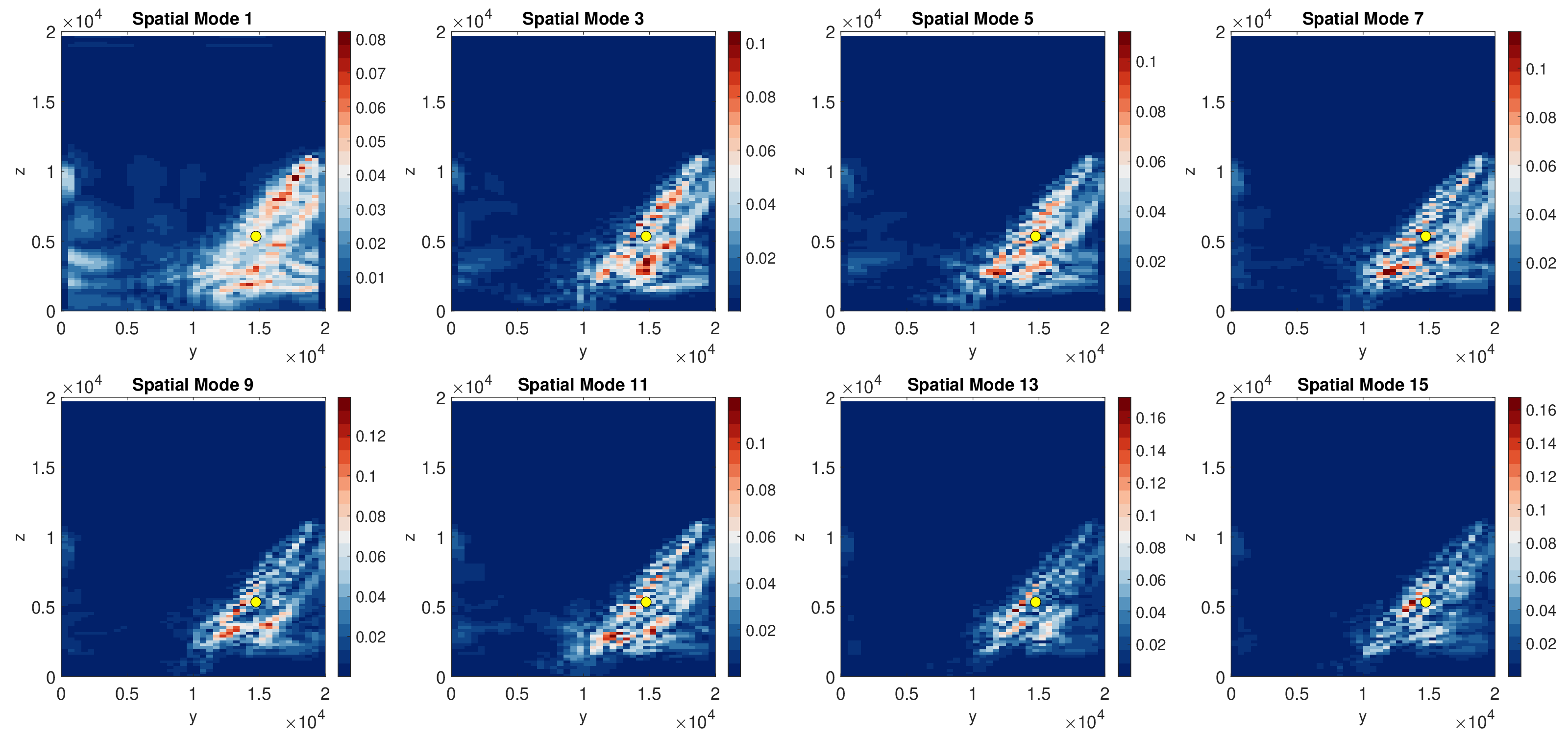}
\caption{Koopman spatial modes selected with SPDMD: Case 2. The absolute values of the Koopman modes $\bm{\phi}_j$ are plotted, with the specific spatial location $(15{\rm{km}}, 7.8{\rm{km}})$ marked by a yellow point.}
\label{fig:vorticity-spatial-modes-V60}
\end{figure}

Since we take the mid-term duration for the dynamic data, which is relatively less transient, most of the eigenvalues have magnitudes close to or even exceeding one (i.e., $|\lambda_i|\geq1$), by comparison with Case~1 in Table~\ref{tabel:vortex-mode-norm-periodic-v30}. 
Fig.~\ref{fig:temporal-vorticity-V60} displays the normal temporal evolutions \eqref{eq:normal-evolution}, exhibiting growing oscillations for modes with $|\lambda_i| > 1$ for $i = 3, 5, \ldots, 15$, except for the first mode with $|\lambda_1| \approx 1$ and zero imaginary part, which reflects a constant, non-oscillatory behavior over time, representing a persistent background vorticity pattern that sets the stage for the transient dynamics associated with heavy rainfall. 
Correspondingly, Fig.~\ref{fig:vorticity-spatial-modes-V60} visualizes the element-wise absolute values of the spatial patterns $|\bm\phi_j|$ for the first $8$ dominant transient modes selected by SPDMD.
Compared to the original data in Fig.~\ref{fig:Ex-1-Vorticity-case-original-data}, most of these spatial modes capture the dominant upwind motion of the bubble-like structural dynamics, representing transient growth patterns that suggest the onset of heavy rainfall. 
The spatial and temporal quantification suggests that each mode, except for the first one, might be utilized for the low-dimensional representation.

\begin{figure}[t!]
\centering 
\includegraphics[width=0.65\linewidth]{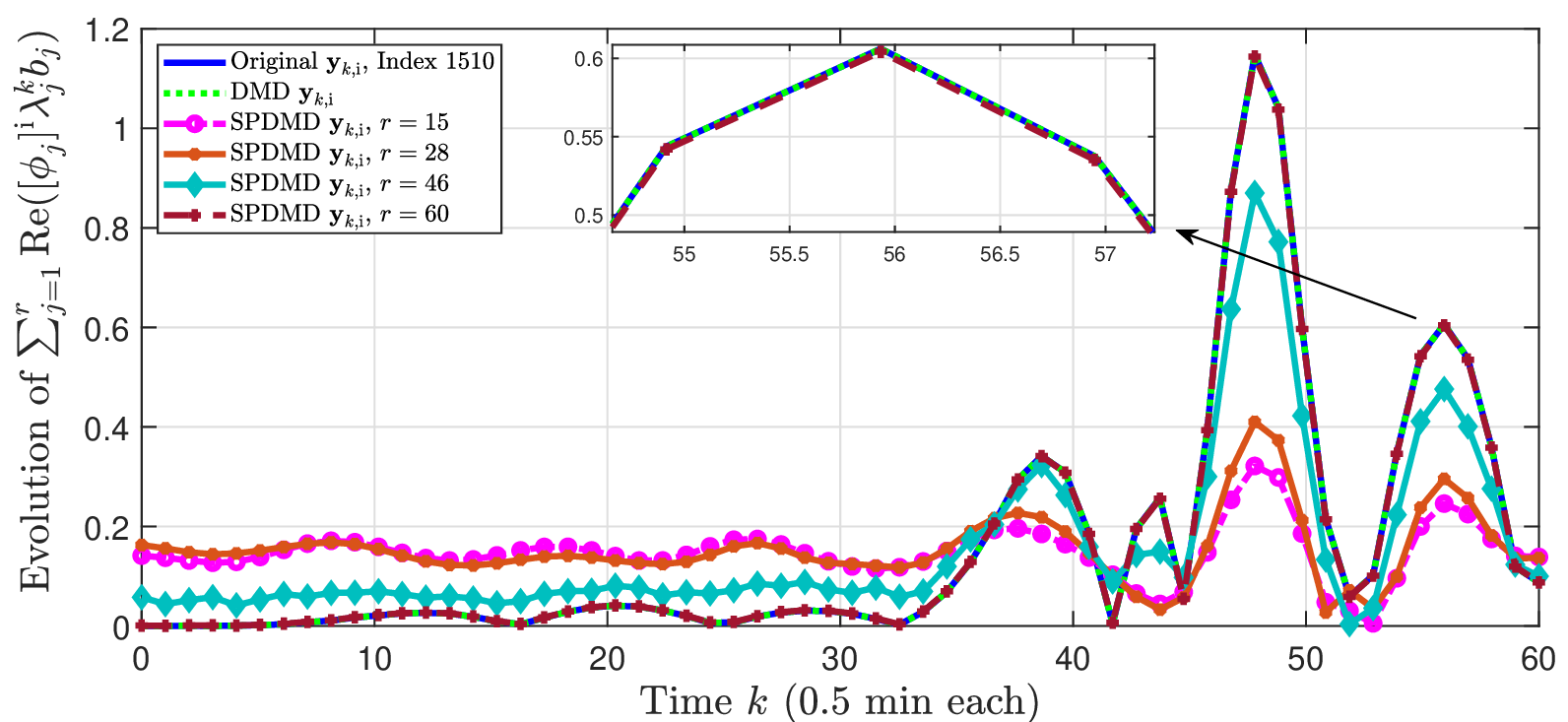}
\includegraphics[width=0.65\linewidth]{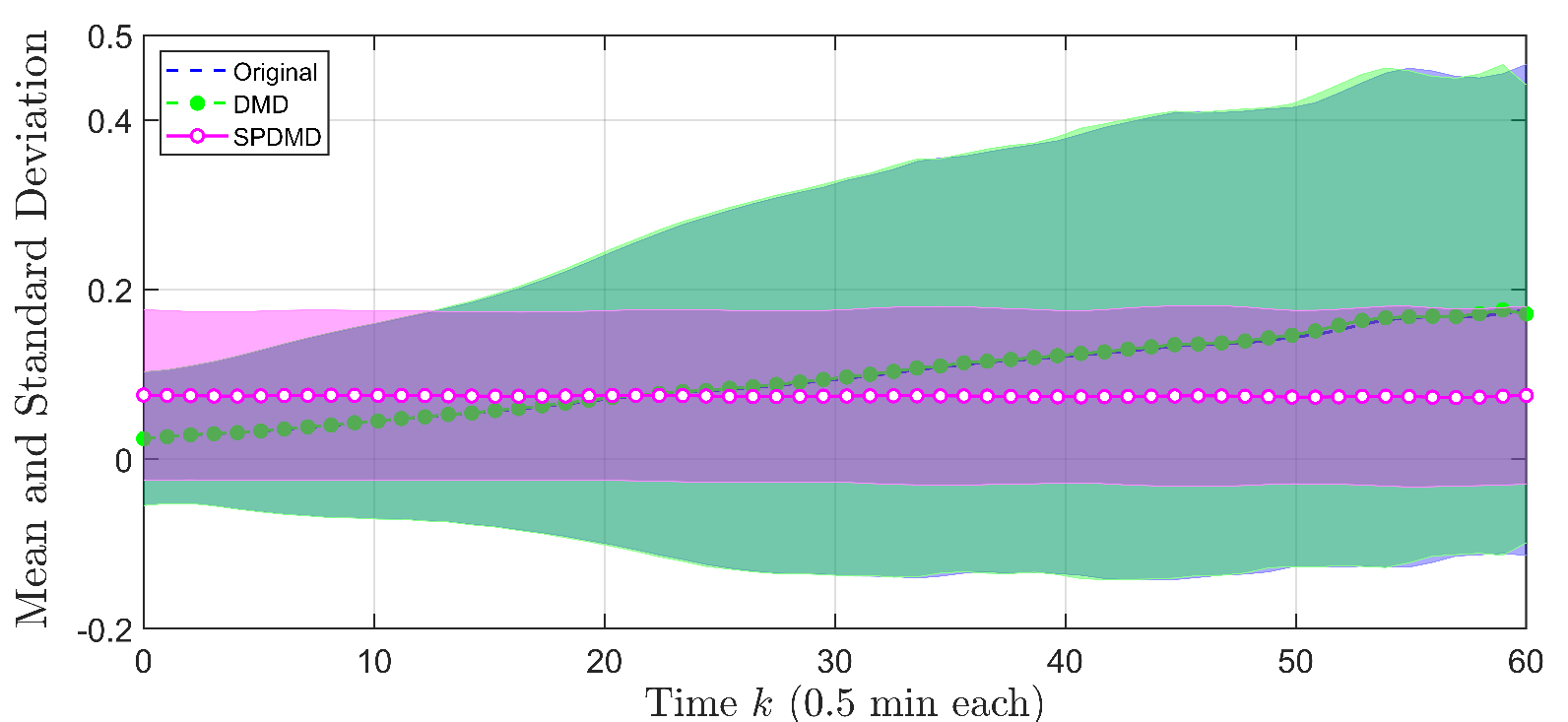}
\caption{(top): Superposition comparison~\eqref{eq:specific-loaction-superposition} of the original vorticity magnitude data, full-mode DMD, and SPDMD method with varying numbers of modes ($r = 15, 28, 46, 60$) at a highlight index $\mathbf{y}_{k,1510}$ over time. Both full-mode DMD (green dashed line) and SPDMD with $r=60$ modes (wine-red dashed line with plus markers) accurately capture the sharp jumps and peaks, whereas reconstructions with fewer modes ($r=15$ or $r=28$) fail to reproduce these jump events. 
(bottom): MSD comparison~\eqref{eq:mean-standard-der} between the original data, full-mode DMD, and SPDMD with selected modes $r = 15$ across all data points in the SCALE simulations.}
 \label{fig:superposition-MSD-recon-vorticity-60}
 \end{figure}

Without loss generality, we focus on the specific location $(y_{30}, z_{38})$, located at approximately $(15~\mathrm{km}, 7.8~\mathrm{km})$ in the SCALE weather domain, as marked by the ``yellow point" in Fig.~\ref{fig:vorticity-spatial-modes-V60}.
Returning to the superposition evolution
$\bfy_{k,\mathrm{i}}\approx\sum_{j=1}^{r}\mathrm{Re}\big([\bm\phi_j]^{\mathrm{i}}\lambda_j^{k}b_j\big)$ defined in \eqref{eq:specific-loaction-superposition},
this spatial location corresponds to the \emph{highlight index} $\mathrm{i}=1510$, where $r$ is the number of selected modes, determined by the sparsity weight $\gamma$.
The comparisons of temporal reconstructions (Original, DMD, and SPDMD with different $r$ modes) at this location are shown in Fig.~\ref{fig:superposition-MSD-recon-vorticity-60}.

\begin{figure}[th!]
\centering
\includegraphics[width=0.7\linewidth]{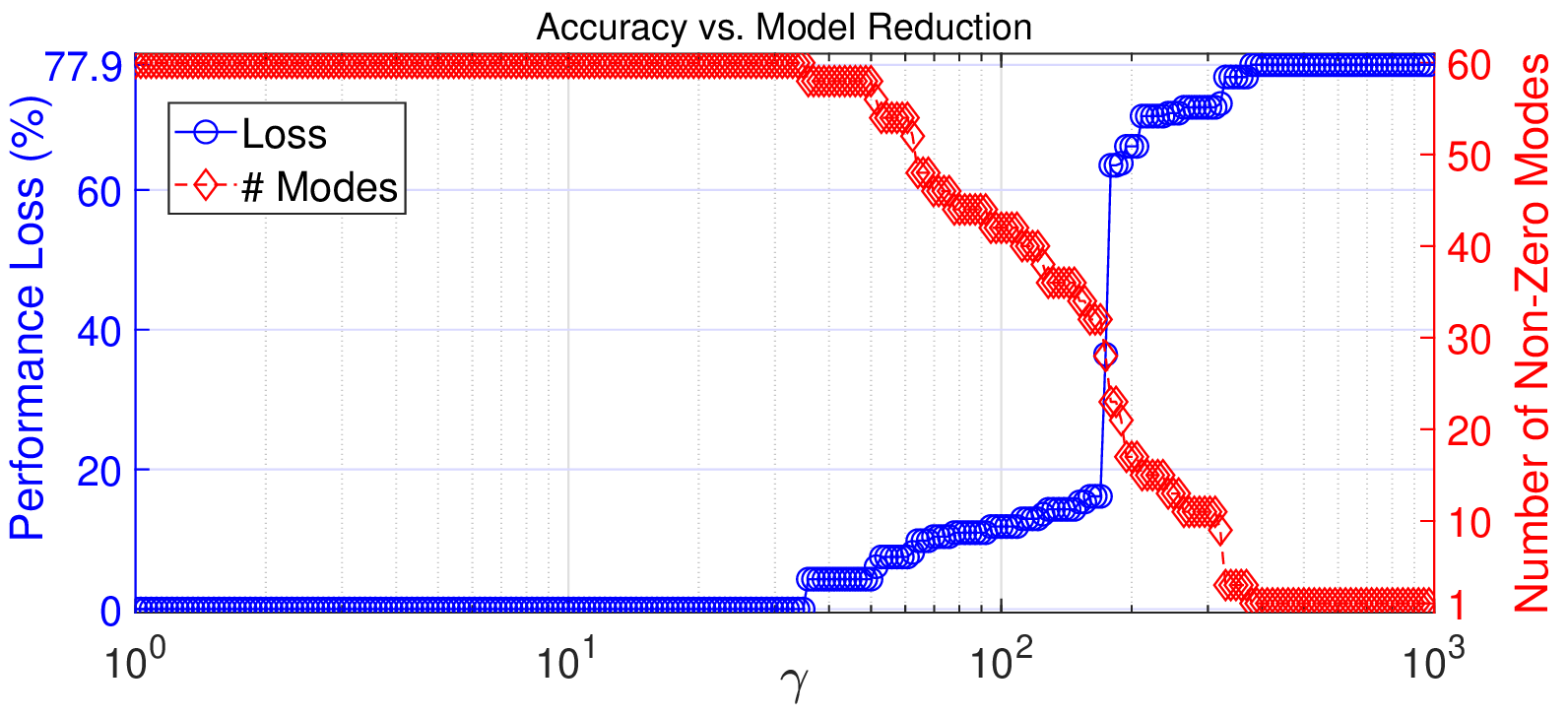}
\caption{The performance loss and the number of non-zero modes governed by the sparsity weight $\gamma\in[1,1000]$ with $250$ grids.}
\label{fig:loss-accuract-vortex-60}
\end{figure} 

The SPDMD method also has limitations, particularly when only a few Koopman modes are used to construct a low-order surrogate model of the weather dynamics.
As illustrated in Fig.~\ref{fig:superposition-MSD-recon-vorticity-60} (top), the ground truth (i.e., original data) of vorticity magnitude exhibits several prominent peaks or jumps resembling exponential growth during the time interval $k\in[35,60]$, corresponding to the weather simulation time window of $[17.5~\mathrm{min}, 30~\mathrm{min}]$.
These significant peaks or jumps naturally motivate us to investigate whether such features can be well-reproduced or tracked by reduced-order surrogate models, as they may indicate potential \emph{weather anomalies}.

In this experiment, the number of Koopman modes $r$ is related to the snapshot length $N$ and sparsity weight $\gamma$ (or constraint $s$, see Remark~\ref{remark:role-of-sparsity}), since the rank of the data matrix $\bfY$ is limited to $r\leq\min(N,p)=N=60$.
We here need to explore the \textbf{factors} (e.g., the number of modes $r$, sparsity weight $\gamma$ or constraint $s\leq{r}$, and snapshot length $N$) that influence whether these sharp peak or jump events are preserved or lost in the reconstruction.
\begin{itemize}
\item 1) Full modes ($r=60$), no sparsity:
With $r=\min(N,p)=60$, no model reduction is applied and sparsity is inactive. Full-mode DMD and SPDMD both near-perfectly recover all sharp jumps and peaks, as seen in Fig.~\ref{fig:superposition-MSD-recon-vorticity-60}.
\item 2) Medium modes ($r=46$), moderate sparsity:
With an appropriate sparsity weight, SPDMD reduces the active modes $r$ from $60$ to $46$. Major jumps and peaks are partially recovered, but some sharp events are missed because not all dominant amplitudes are retained. Still, the model provides a reasonable reduced-order approximation.
\item 3) Few modes ($r=28,15$), strong sparsity:
Under strong sparsity, many necessary amplitudes are removed. As a result, sharp jumps and peaks are largely lost, leading to clear degradation compared with the full and medium-mode cases.
\end{itemize}
A similar conclusion is further supported by the MDS comparison \eqref{eq:mean-standard-der}, which evaluates reconstruction results across all vorticity data points in the $y$–$z$ plane, as shown in Fig.~\ref{fig:superposition-MSD-recon-vorticity-60} (bottom).
This behavior may result from the sparsity constraint, which suppresses sharp peaks in the amplitude.
As a result, the summation over the selected sparse modes in \eqref{eqn:reconst-SPDMD-amplitudes-order} does not include all large-amplitude contributions, which are ordered by magnitude in~\eqref{eqn:order-amplitudes-b}. Consequently, some sharp peaks or jumps in the temporal evolution are underrepresented or missing. When such sharp jump events require the collective contribution of amplitudes (or modes), discarding these amplitude terms hinders accurate reconstruction, leading to peak or jump loss.

Similar to previous experiments, we also demonstrate the trade-off between reconstruction accuracy and model reduction using SPDMD, as displayed in Fig.~\ref{fig:loss-accuract-vortex-60}.
There is no doubt that increasing the sparsity weight $\gamma$ reduces the number of retained modes, ranging from the full $60$ modes at $\gamma_{\min}=1$ to only a single mode at the largest weight $\gamma_{\max}=1000$, evaluated over $250$ grid points. This sparsification therefore leads to a loss in reconstruction accuracy, with performance degradation increasing from nearly $0\%$ up to $77.9\%$.

\section{Conclusion and Discussion}\label{sec:conclusion-discussion}
The Koopman operator framework is powerful for analyzing short-term weather simulations across various observational data fields using Koopman modes, without imposing predefined mathematical models on the weather system. In this paper, we demonstrated that sparsity-promoting dynamic mode decomposition (SPDMD) effectively extracts transient Koopman modes from finite-length time snapshots of high-dimensional SCALE weather convective simulations. In particular, the leading spatial modes captured through SPDMD exhibit warm bubble-like patterns, offering physical insights (e.g., precipitation and weather anomalies) into the evolution of complex weather phenomena.
Furthermore, by flexibly tuning the sparsity weight, we balance reconstruction accuracy and model complexity, allowing a trade-off between capturing dominant dynamical behaviors and obtaining an effective low-dimensional surrogate model.

However, several limitations remain. When only a few SPDMD-selected modes are retained across different observables, the reconstructed superposition and MSD profiles may largely deviate from the original data and miss sharp jumps or peaks, motivating the following issues (i)–(iii).
\begin{itemize}
\item[(i)] \textbf{Choice of Observables}: 
The performance loss $\Pi\%$ reveals a clear accuracy and model complexity trade-off. 
The velocity observable yields a low loss (up to $13.8\%$), whereas the vorticity magnitude shows a much higher loss (about $76\%$), indicating strong sensitivity to the observable choice. 
This reflects the underlying observable subspace: low loss implies suitable observables and reliable surrogate modeling, while high loss suggests missing sharp jumps or peaks. Other observables (e.g., humidity ratio data field) were examined in \cite{zhang2025KoopmanMode}. 
\item[(ii)] \textbf{Peak and Jump Loss vs.~Lossless}: 
The recovery of sharp peaks and jumps depends on the number of retained modes ($r \le N$). 
Full-mode DMD and SPDMD accurately capture these events, whereas moderate truncation preserves only overall trends, and few modes fail to recover sharp peaks. 
The loss occurs because essential small to moderate amplitudes are discarded. The SPDMD alleviates this issue by tuning the sparsity weight to balance accuracy and complexity, though the trade-off ultimately depends on data quality and observable choice.
\item[(iii)] \textbf{Point vs.~Continuous Spectrum}: 
The DMD and SPDMD capture the point spectrum of Koopman operator. 
In general, the decomposition of nonlinear time series in terms of spectral properties of the Koopman operator also contains a continuous spectrum \cite{mezic2005spectral,colbrook2024rigorous}, associated with mixing, noise, non-normal transients, and chaos.  In finite-rank DMD, these effects appear as many weak modes; sparse or heavily truncated reconstructions suppress them, which may contribute to peak loss.
 \end{itemize}
 
Future work will investigate multiscale weather dynamics by separating slow and fast components using iterative, multi-resolution DMD (mrDMD) method \cite[Chap.~5]{kutz2016dynamic} on weather data fields across transient stages. 
Another direction is to examine continuous-spectrum effects in reduced-order transient dynamics via Residual DMD \cite{colbrook2024rigorous}.

\section*{Acknowledgment}
 This work was supported part by JST Moonshot R\&D Grant Number JPMJMS2284.
This work used computational resources of Fugaku provided by RIKEN through the HPCI System Research Project (Project ID: hp240169, hp250178).

\section*{Data Availability}
Partially selected snapshots of the original data are shown in Figs.~\ref{fig:Ex-0-velocity-case-original-data} and~\ref{fig:Ex-1-Vorticity-case-original-data}.\\
Reanalysis and observed data were obtained from SCALE Weather Simulation software (\url{https://scale.riken.jp}), along with the associated \href{https://drive.google.com/drive/folders/1aIjP4sDFhO6NG296vyf7FubWPyPo0VtE?usp=drive_link}{Data Sources} and a publicly accessible code repository on Github:\\~\url{https://github.com/zc-zhang/Transient-Weather-Modes-SPDMD}. \\



\section*{Appendix}
\begin{algorithm}[H]
\caption{Sparsity-Promoting DMD (SPDMD Algorithm) \cite{jovanovic2014sparsity}}
\label{algo:spdmd}
\begin{algorithmic}[1]
\Require Snapshot matrix \text{Data}, sparse weight $\gamma$ with $\gamma_{\min}$, $\gamma_{\max}$, $\gamma_{\texttt{grids}}$, parameters $\rho$, snapshot $N$, convergence tolerances $\epsilon_{\text{primal}}$, $\epsilon_{\text{dual}}$, ADMM iterations $k_{\text{max}}$
\Ensure Sparse amplitudes $\bfb_{r}$, Koopman modes ${\bm\Phi}_r$, Performance losses ${J}_{\text{sp}}$, ${J}_{\text{pol}}$, ${J}_{\text{loss}}$
\State $\bfY \gets \text{Data}(:, 1:N-1)$, $\bfY' \gets \text{Data}(:, 2:N)$ 
\Comment{Split data in Eq.~\eqref{eq:separate-data-Y}}
\State $\bfU$, $\bmSigma$, ${\bfV}^{\ast} \gets \mathtt{svd}(\bfY)$ \Comment{Perform SVD}
\State $r \gets \mathrm{rank}(\bmSigma)$ \Comment{Determine rank}
\State ${\bfU}_r \gets {\bfU}(:, 1:r)$, ${\bmSigma}_r \gets {\bmSigma}(1:r, 1:r)$, ${\bfV}_r \gets {\bfV}(:, 1:r)$ 
\Comment{Truncate}
\State $\tilde{\bfA} \gets \bfU_{r}^{\ast} \bfY^{\prime} \bfV_{r} \bmSigma_{r}^{-1}$ 
\Comment{Reduced-order matrix in Eq.~\eqref{eq:decomposition-tilde_A}}
\State $\bfW$, $\bmLambda \gets \mathtt{eig}\big(\tilde{\bfA}\big)$ \Comment{Eigenvalues, eigenvectors}
\State ${\bm\Phi}_r \gets {\bfU}_r \bfW$ 
\Comment{DMD modes}
\For{$i = 1$ to $N$} \Comment{Vandermonde matrix}
    \State $\bfT_r(:, i) \gets \operatorname{diag}(\bmLambda)^{i-1}$ 
\EndFor
\State $\mathbf{L} \gets {\bfU}_r\bfW, \mathbf{R} \gets \bfT_r, \mathbf{G} \gets {\bmSigma}_r{\bfV}_{r}^\top$ 
\Comment{Optimization variablies}
\State $\mathbf{P} \gets \mathbf{L}^{\top} \mathbf{L} \cdot \mathtt{conj}(\mathbf{R} \mathbf{R}^{\top})$, $\mathbf{d} \gets \operatorname{conj}\big(\operatorname{diag}(\mathbf{R} \mathbf{G}^{\top} \mathbf{L})\big)$, $\eta \gets \operatorname{trace}(\mathbf{G}^{\top} \mathbf{G})$ 
\Comment{QP setup in Eq.~\eqref{eq:QP-closed-form-solution}}
\State $\mathbf{P}_{\mathrm{chol}} \gets \operatorname{chol}(\mathbf{P})$ 
\Comment{Cholesky factorization}
\State $\bfb_{\text{init}} \gets \mathbf{P}_{\mathrm{chol}}^{-\top}\big(\mathbf{P}_{\mathrm{chol}}^{-1} \mathbf{d}\big)$ 
\Comment{Non-sparse amplitudes}
\State ${\bm\theta} \gets 0$, ${\bm\xi} \gets 0$, $k \gets 0$ 
\Comment{Initialize ADMM}
\State $\hat{\bm\xi}_{\text{prev}} \gets 0$ \Comment{Store previous $\hat{\bm\xi}$ for dual residual}
\While{$k < k_{\text{max}}$ and not converged} \Comment{Sparse optimization via ADMM}
    \State $\widehat{\bfb} \gets \left(\mathbf{L}^\top \mathbf{L} + \rho \bfI\right)^{-1} \left(\mathbf{L}^\top \mathbf{G} + \rho {\bm\xi} - {\bm\theta}\right)$ \Comment{Update amplitudes $\bfb$}
    \State $\hat{\bm\xi} \gets \operatorname{shrinkage}\left(\widehat{\bfb} + \frac{1}{\rho}{\bm\theta}, \frac{\gamma}{\rho}\right)$ 
    \Comment{Soft-thresholding for sparsity}
    \State ${\bm\theta} \gets {\bm\theta} + \rho \left(\widehat{\bfb} - \hat{\bm\xi}\right)$ 
    \State Check $\left\|\widehat{\bfb} - \hat{\bm\xi}\right\|_2 \leq \epsilon_{\text{primal}}$, $\left\|\rho \left(\hat{\bm\xi} - \hat{\bm\xi}_{\text{prev}}\right)\right\|_2 \leq \epsilon_{\text{dual}}$ 
    \State $\hat{\bm\xi}_{\text{prev}} \gets \hat{\mathbf{z}}$, $k \gets k + 1$ 
\EndWhile
\State $\bfb_{r} \gets \hat{\bm\xi}$ 
\Comment{Set sparse amplitudes}
\State $\mathcal{S}^{c} \gets \operatorname{find}\big(\hat{\bm\xi} = 0\big)$ \Comment{Identify zero amplitudes}
\State Solve $\min_{\bfb_{\text{pol}}} \|\mathbf{G} - \mathbf{L} \operatorname{diag}(\bfb_{\text{pol}}) \mathbf{R}\|_{\text{F}}^2$ over non-zero indices in $\mathcal{S}^{c}$ \Comment{Polishing step}
\State ${J}_{\text{sp}} \gets \|\mathbf{G} - \mathrm{L} \operatorname{diag}(\bfb_r) \mathbf{R}\|_{\text{F}}^2$ \Comment{Performance losses}
\State ${J}_{\text{pol}} \gets \|\mathbf{G} - \mathbf{L} \operatorname{diag}(\bfb_{\text{pol}}) \mathbf{R}\|_{\text{F}}^2$
\State $\widehat{\mathbf{G}} \gets \operatorname{trace}(\mathbf{G}^\top \mathbf{G})$
\State ${J}_{\text{loss}} \gets 100 \sqrt{{J}_{\text{sp}} / \widehat{\mathbf{G}}}$ \Comment{Relative loss in percentage}

\State \Return $\bfb_r$, $\bm\Phi_r$, ${J}_{\text{sp}}$, ${J}_{\text{pol}}$, ${J}_{\text{loss}}$
\end{algorithmic}
\end{algorithm}

\bibliographystyle{IEEEtran}  
\bibliography{main}  

\clearpage
\renewcommand{\thefigure}{A\arabic{figure}}
\setcounter{figure}{0} 

\begin{figure}
    \centering
        \includegraphics[width=0.9\textwidth,keepaspectratio]{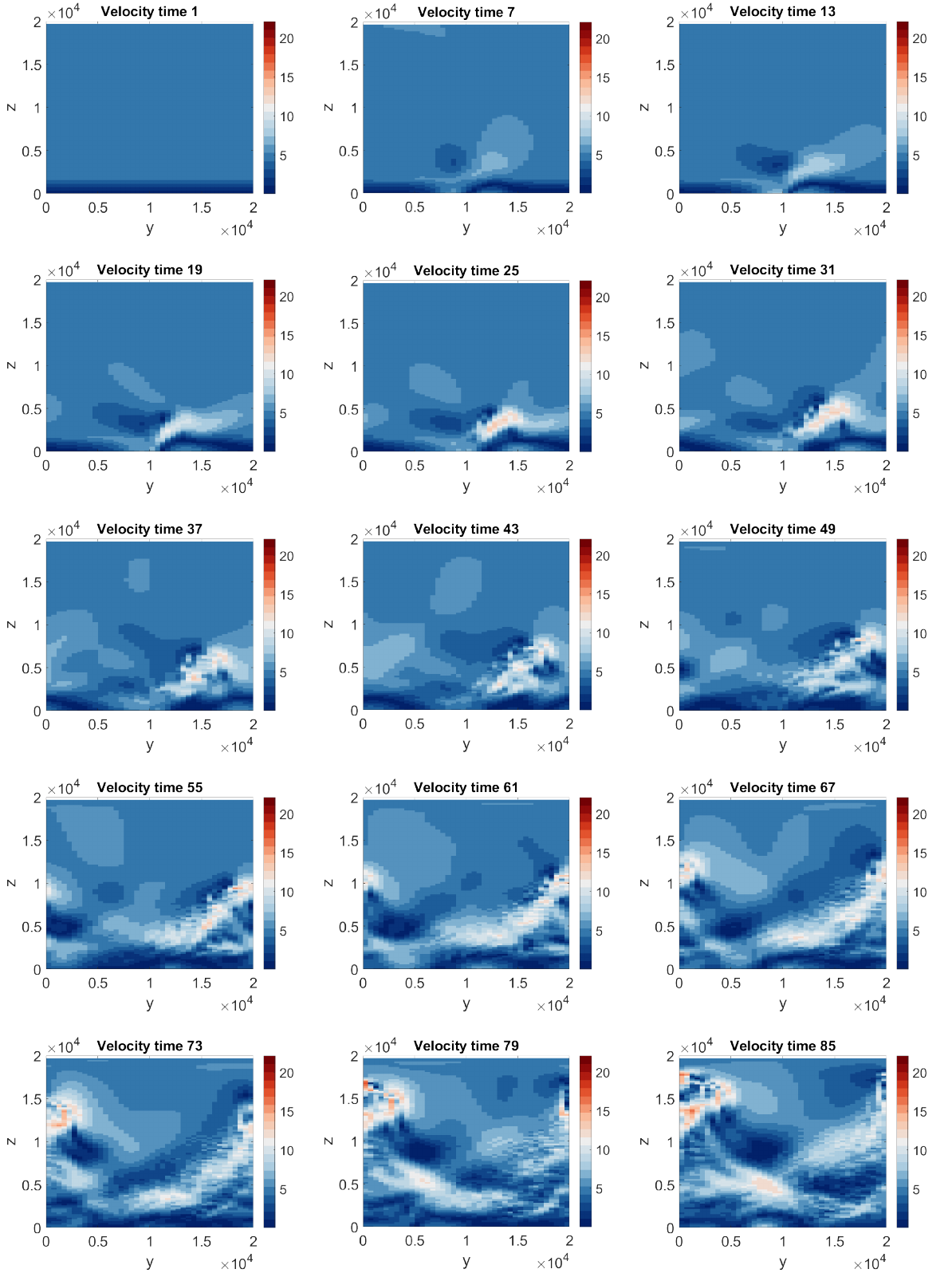}
    \caption{Original data of velocity magnitude: Partially selected snapshots from the evolution of the velocity magnitude field exhibit warm bubble-like patterns rising upward.}
    \label{fig:Ex-0-velocity-case-original-data}
\end{figure}

\begin{figure}
    \centering
       \includegraphics[width=0.9\textwidth,keepaspectratio]{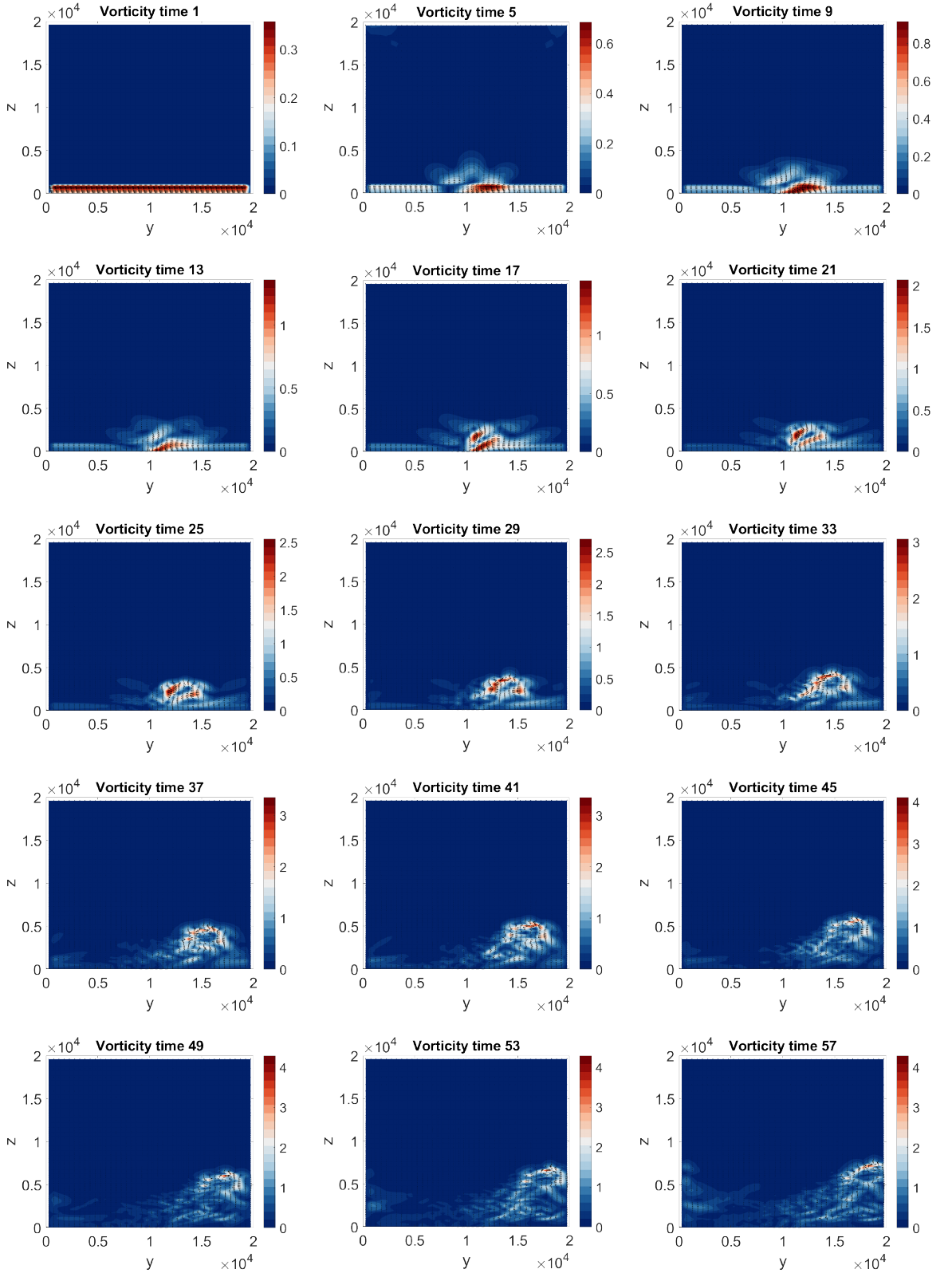}
 \caption{Original data of vorticity magnitude: Partially selected snapshots from Cases 1 and 2 show the evolution of the vorticity magnitude field, revealing the emergence and growth of warm bubble-like patterns.}
    \label{fig:Ex-1-Vorticity-case-original-data}
\end{figure}

\end{document}